\documentclass[11pt,prd,a4paper,preprintnumbers,amsmath,amssymb,nofootinbib]{article}
\pdfoutput=1
\usepackage{feynmp-auto,expdlist}
\usepackage{amsmath,amsfonts,amssymb}
\usepackage{graphicx}
\usepackage{enumerate}
\usepackage{hyperref}
\usepackage{axodraw2}
\usepackage{latexsym}
\usepackage{hepnicenames}
\usepackage{enumerate}
\usepackage{soul}
\usepackage[normalem]{ulem}
\usepackage{wasysym}
\usepackage{makecell}
\usepackage{bbm}

\font\tenrsfs=rsfs10 at 12pt
\font\sevenrsfs=rsfs7
\font\fiversfs=rsfs5
\newfam\rsfsfam
\textfont\rsfsfam=\tenrsfs
\scriptfont\rsfsfam=\sevenrsfs
\scriptscriptfont\rsfsfam=\fiversfs

\numberwithin{equation}{section}

\usepackage{mathrsfs}
\usepackage{braket}
\usepackage{titling}
\usepackage{amsmath}
\usepackage{slashed}
\usepackage{amssymb}
\usepackage{epsfig}
\usepackage{graphicx}
\usepackage{color}
\usepackage{rotating}
\usepackage{hyperref}
\usepackage[margin=1.in]{geometry}
\usepackage[table,xcdraw,dvipsnames]{xcolor}
\usepackage[compress,numbers,sort]{natbib}
\usepackage{colortbl}
\usepackage{pdflscape}
\usepackage{color}
\usepackage{mathtools}
\usepackage{colortbl}
\usepackage{comment}
\definecolor{Gray}{gray}{0.95}
\definecolor{RGray}{gray}{0.85}
\definecolor{CGray}{gray}{0.93}
\definecolor{piggypink}{rgb}{0.99, 0.87, 0.9}
\definecolor{babyblue}{rgb}{.67,.83,.99}

\newcommand{\X}{{\cal X}}

\newcommand{\SU}{{\rm SU}}

\newcommand{\U}{{\rm U}}

\renewcommand{\L}{{\cal L}}
\newcommand{\N}{{\cal N}}
\newcommand{\E}{{\cal E}}
\newcommand{\Y}{{\cal Y}}

\definecolor{nicered}{rgb}{0.7,0.1,0.1}
\definecolor{nicegreen}{rgb}{0.1,0.5,0.1}
\definecolor{red}{rgb}{1.0, 0, 0}
\definecolor{niceblue}{rgb}{0,0,0.8}
\definecolor{red}{rgb}{1.0, 0, 0}
\hypersetup{colorlinks,bookmarksopen,bookmarksnumbered,
linkcolor=blus,pdfstartview=FitH,urlcolor=rossos,citecolor=verde}
\allowdisplaybreaks

\definecolor{rosso}{cmyk}{0,1,1,0.4}
\definecolor{rossos}{cmyk}{0,1,1,0.55}
\definecolor{rossoc}{cmyk}{0,1,1,0.2}
\definecolor{blu}{cmyk}{1,1,0,0.3}
\definecolor{blus}{cmyk}{1,1,0,0.6}
\definecolor{bluc}{cmyk}{1,1,0,0.1}
\definecolor{verde}{cmyk}{0.92,0,0.59,0.25}
\definecolor{verdec}{cmyk}{0.92,0,0.59,0.15}
\definecolor{verdes}{cmyk}{0.92,0,0.59,0.4}

\def\eq#1{{Eq.~(\ref{#1})}}
\def\eqs#1#2{{Eqs.~(\ref{#1})--(\ref{#2})}}
\def\fig#1{{Fig.~\ref{#1}}}

\def\Table#1{{Table~\ref{#1}}}

\def\sect#1{{Sect.~\ref{#1}}}

\def\app#1{{App.~\ref{#1}}}

\def\vev#1{\left\langle #1\right\rangle}

\renewcommand{\bar}{\overline}

\newcommand{\beq}{\begin{equation}}
\newcommand{\eeq}{\end{equation}}
\newcommand{\bea}{\begin{eqnarray}}
\newcommand{\eea}{\end{eqnarray}}

\renewcommand{\[}{\left[}
\renewcommand{\]}{\right]}
\renewcommand{\(}{\left(}
\renewcommand{\)}{\right)}

\renewcommand{\S}{\mathcal{S}}
\renewcommand{\X}{\mathcal{X}}

\def\cA{{\cal A}}
\def\cL{{\cal L}}
\def\be{\begin{equation}}
\def\ee{\end{equation}}

\begin{document}



\begin{center}  
{\LARGE
\bf\color{blus} 
Light vectors coupled to anomalous currents 
\\ \vspace{0.3cm} 
with harmless Wess-Zumino terms
} \\
\vspace{0.8cm}

{\bf Luca Di Luzio$^{a,b}$, Marco Nardecchia$^{c}$, Claudio Toni$^{c}$ }\\[7mm]

{\it $^a$Dipartimento di Fisica e Astronomia `G.~Galilei', Universit\`a di Padova, Italy}\\[1mm]
{\it $^b$Istituto Nazionale Fisica Nucleare, Sezione di Padova, Italy}\\[1mm]
{\it $^c$Physics Department and INFN Sezione di Roma La Sapienza, \\ 
Piazzale Aldo Moro 5, 00185 Roma, Italy}\\[1mm]

\vspace{0.3cm}
\begin{quote}
We reconsider the case of light vectors coupled to anomalous fermionic currents, 
focussing on the interplay between UV and IR dynamics. 
Taking as a general framework the gauging of the Standard Model accidental symmetries, 
we show that it is possible to devise an anomaly-free UV completion 
with mostly chiral heavy fermions 
such that anomalous Wess-Zumino 
terms are 
suppressed in the IR,
thus relaxing would-be strong 
bounds 
from the longitudinal emission 
of light 
vectors 
coupled to non-conserved currents. 
We classify such scenarios and 
show that they will be 
extensively 
probed at the high-luminosity phase of the LHC 
via the measurement of the $h \to Z \gamma$ rate 
and the direct search for non-decoupling charged leptons. 

\end{quote}
\thispagestyle{empty}
\end{center}

\bigskip
\tableofcontents


\section{Introduction}
\label{sec:intro}

The physics of light 
spin-1 dark bosons 
has witnessed a growing amount of interest in the recent years, 
from both a theoretical and phenomenological standpoint.   
A standard benchmark 
is that of a 
secluded $\U(1)$ gauge boson, kinetically mixed with the photon \cite{Holdom:1985ag} and hence 
universally coupled to the Standard Model (SM) sector via the electromagnetic current. 
This framework, although elegant and predictive, can be 
too restrictive for phenomenological applications 
and hence more general forms of the light vector boson interactions with the SM fields can be envisaged.  
Going beyond the kinetic mixing framework, a theoretically motivated 
option is provided by the gauging of the accidental global symmetries of the SM,  
that is baryon number $\U(1)_B$ and family lepton number $\U(1)_{L_i}$ (with $i = e,\mu,\tau$) 
in the limit of massless neutrinos.  
Within the SM field content, only the $L_i-L_j$ combinations turn out to be anomaly free \cite{Foot:1990mn,He:1990pn,He:1991qd}. Hence, 
in order to consistently gauge a general linear combination 
\beq
\label{eq:genU1X}
X = \alpha_B B + \sum_{i=e,\, \mu,\, \tau} \alpha_i L_i \, ,
\eeq 
one requires 
new fermions, also known as \emph{anomalons},
which cancel the anomalies of the new $\U(1)_X$ factor, 
also  
in combination 
with the electroweak gauge group. 
Note that \eq{eq:genU1X} is the most general linear combination of abelian global symmetries of the SM that 
can be gauged, 
under the assumption that all SM Yukawa operators are allowed in the quark sector at the renormalizable level. 
Barring the cases of $B/3 - L_i$ and linear combinations thereof,  
the anomalons 
need to be charged under the electroweak gauge group (henceforth indicated more precisely as electroweak anomalons).\footnote{This is not the case if 
some of the quark Yukawa operators arise at the non-renormalizable level, 
as e.g.~discussed recently in Ref.~\cite{Greljo:2021npi}.}     
In the latter case, in order to 
evade detection at high-energy particle colliders,  
the new fermions need to be heavier than the electroweak scale. 
Consequently, their effects on the physics of the light vector boson associated with the $\U(1)_X$ gauge symmetry, 
here denoted as $\X$, 
can be described within 
an effective field theory (EFT) approch. In particular, 
after integrating out the new heavy fermions at one loop, 
one generates 
dimension-4 Wess-Zumino (WZ) 
terms, schematically 
of the form $\X (W \partial W + WWW)$ and $\X B \partial B$ 
(with $W$ and $B$ denoting $\SU(2)_L$ and $\U(1)_Y$ gauge bosons).  
These 
contact interactions have the role of compensating, 
in the EFT 
without electroweak anomalons,  
the 
anomalous shift of the effective action 
due to the anomalous SM fermion current coupled to $\X$ 
(see e.g.~\cite{DHoker:1984izu,DHoker:1984mif,Preskill:1990fr,Feruglio:1992fp}).  

As it was emphasized more recently in Refs.~\cite{Dror:2017ehi,Dror:2017nsg}, 
WZ terms 
display an axion-like behaviour 
(as can be understood by applying the equivalence theorem to the longitudinal component of $\X$)
and lead to amplitudes that grow with the energy. 
The anomalous $\X W \partial W$ vertex can be dressed with SM flavour-violating 
interactions leading to loop-induced flavour changing neutral current (FCNC) processes,  
while the anomalous $\X B \partial B$ vertex is responsible for  
$Z \to \gamma \X$ decays at the tree level (see also \cite{Ismail:2017fgq,Michaels:2020fzj,Davighi:2021oel,Kribs:2022gri}). 
In both cases these processes 
are enhanced as $(\text{energy} / m_{\X})^2$, thus 
resulting into the typically most stringent bounds on light vectors 
with no direct couplings to electrons, 
as e.g.~in the case of gauged baryon number. 

It is known (see e.g.~\cite{Dror:2017nsg})
that in the limit where the mass of the anomalons 
stems from a SM-preserving vacuum expectation value (VEV), 
the low-energy coefficients of the WZ terms are entirely 
fixed by the requirement of cancelling the $\SU(2)_L^2\U(1)_X$ and $\U(1)_Y^2\U(1)_X$ anomalies of the SM  
sector. 
On the other hand, if the anomalons pick up a mass contribution from the electroweak VEV then 
the coefficients of the WZ terms become model-dependent. In particular, in the limit where the anomalons mass is completely 
due to electroweak symmetry breaking sources, the anomalous couplings of the longitudinal component of  
$\X$ with SM electroweak gauge bosons goes to zero, 
thus relaxing the above mentioned strong bounds on light 
vectors. 

In this paper, we revisit the argument why WZ terms become harmless in the limit where the 
electroweak anomalons obtain their mass solely from the Higgs,  
classify the structure of UV completions that allow for such a pattern 
and discuss their electroweak-scale phenomenology. 
Due to its non-decoupling nature, the phenomenology of the 
electroweak anomalons is tightly constrained 
(but not yet ruled out) 
by Higgs couplings measurements and direct searches, thus making the whole
setup testable at the high-luminosity phase of the LHC (HL-LHC).  
Chiral fermionic extensions of the SM, sharing some similarities with our setup, 
were previously discussed in a different context in Refs.~\cite{Bizot:2015zaa,Bonnefoy:2020gyh}.
Here, the main phenomenological interest 
consists in the physics of light 
(i.e.~sub-GeV) vector bosons coupled to anomalous SM currents 
and the possibility of 
re-opening a large portion of parameter space, which 
might be probed by 
several
low-energy experiments or help in explaining current experimental anomalies, 
such as e.g.~that of the muon $g$-2 \cite{Muong-2:2021ojo}.  

The paper is structured as follows. \sect{sec:generalU1} is the core of the work, 
in which we provide the general setup for the gauging of the generic 
linear combination of U(1) factors in \eq{eq:genU1X}.  
We discuss in particular the heavy anomalons sector leading to the 
cancellation of gauge anomalies 
and compute 
the resulting WZ terms in the EFT. 
In passing, we also deal with the issue of neutrino masses 
when lepton family generators are gauged. 
\sect{sec:anomalonspheno} is devoted instead to the phenomenology 
of the electroweak anomalons, 
in the limit where their mass dominantly stems from the Higgs VEV. 
We conclude in \sect{sec:conclusions}, while in 
\app{app:WZterms} we collect a series of technical results about the calculation of WZ terms. 

\section{Gauging the Standard Model accidental symmetries}
\label{sec:generalU1}

In this Section, we provide an explicit UV completion for the gauging of the most general combination 
of the SM global symmetries in \eq{eq:genU1X}, discuss the 
the conditions for the cancellation of gauge anomalies,  
compute the 
spectrum and the EFT below the scale 
of the heavy fermions (anomalons) 
assuming that the only new physics light state is the vector boson $\X$.

\subsection{UV model}

The field content of the model is displayed in 
\Table{tab:genU1fieldcontent}, where the anomalon fields are highlighted in color and we also extended 
the scalar sector of the SM in order to spontaneously break the $\U(1)_X$ symmetry. 
Similar setups for anomaly cancellation 
were considered e.g.~in Refs.~\cite{Duerr:2013dza,Duerr:2013lka,Dobrescu:2014fca,Dobrescu:2015asa,FileviezPerez:2019jju,FileviezPerez:2019cyn}.\footnote{For other anomalon configurations leading to anomaly cancellation 
when baryon and/or lepton number generators are gauged  
see e.g.~\cite{FileviezPerez:2011pt,FileviezPerez:2014lnj}.} 
Here, 
the more general SM charges of the electroweak anomalon fields $(\L,\N,\E)$ are needed 
to evade LHC constraints on purely-chiral fermions 
for $\Y \approx 2,-1$ \cite{Bizot:2015zaa,Bonnefoy:2020gyh}, as it will be reviewed in \sect{sec:anomalonspheno}. 
In fact, as already anticipated in the Introduction, 
we will be interested in exploring the limit in which the electroweak anomalon masses are dominantly 
due to the Higgs, so that the strong bounds stemming from the 
anomalous WZ 
couplings of the light vector with SM gauge bosons are relaxed.
We have also included $N$ copies of chiral SM-singlet fermions $\nu^{\alpha}_{R}$ 
($\alpha = 1, \ldots, N$)
which allow to have more freedom for the cancellation of 
$\U(1)_X$ and $\U(1)^3_X$ anomalies (as well as provide a seesaw setup for neutrino masses), 
but whose presence does not impact the calculation of the electroweak WZ terms.

\begin{table}[!ht]
	\centering
	\begin{tabular}{|c|c|c|c|c|c|}
	\rowcolor{CGray} 
	\hline
	Field & Lorentz & $\SU(3)_C$ & $\SU(2)_L$ & $\U(1)_Y$ & $\U(1)_{X}$ \\ 
	\hline 
	$q^{i}_L$ & $(\tfrac{1}{2},0)$ & 3 & 2 & 1/6 & $\alpha_B/3$ \\ 
	$u^{i}_R$ & $(0,\tfrac{1}{2})$ & 3 & 1 & 2/3 & $\alpha_B/3$ \\ 
	$d^{i}_R$ & $(0,\tfrac{1}{2})$ & 3 & 1 & $-1/3$ & $\alpha_B/3$ \\ 
	$\ell^{i}_L$ & $(\tfrac{1}{2},0)$ & 1 & 2 & $-1/2$ & $\alpha_i$ \\ 
	$e^{i}_R$ & $(0,\tfrac{1}{2})$ & 1 & 1 & $-1$ & $\alpha_i$ \\ 
	$H$ & $(0,0)$ & 1 & 2 & 1/2 & $0$ \\
	\hline
	\rowcolor{piggypink} 
	$\L_L$ & $(\tfrac{1}{2},0)$ & 1 & 2 & $\Y-1/2$ & $X_{\L_L}$ \\
	\rowcolor{piggypink} 
	$\L_R$ & $(0,\tfrac{1}{2})$ & 1 & 2 & $\Y-1/2$ & $X_{\L_R}$ \\
	\rowcolor{piggypink} 
	$\E_L$ & $(\tfrac{1}{2},0)$ & 1 & 1 & $\Y-1$ & $X_{\E_L}$ \\
	\rowcolor{piggypink} 
	$\E_R$ & $(0,\tfrac{1}{2})$ & 1 & 1 & $\Y-1$ & $X_{\E_R}$ \\
	\rowcolor{piggypink} 
	$\N_L$ & $(\tfrac{1}{2},0)$ & 1 & 1 & $\Y$ & $X_{\N_L}$ \\
	\rowcolor{piggypink} 
	$\N_R$ & $(0,\tfrac{1}{2})$ & 1 & 1 & $\Y$ & $X_{\N_R}$ \\ 
	\rowcolor{piggypink} 
	$\nu_R^{\alpha}$ & $(0,\tfrac{1}{2})$ & 1 & 1 & $0$ & $X^{\alpha}_{\nu_R}$ \\
	$\S$ & $(0,0)$ & 1 & 1 & 0 & $X_{\S}$ \\  
	\hline
	\end{tabular}	
	\caption{\label{tab:genU1fieldcontent} 
	Anomaly-free field content for a general $\SU(3)_C \times \SU(2)_L \times \U(1)_Y \times \U(1)_X$ gauge theory, 
	with $X = \alpha_B B + \sum_{i=e,\mu,\tau} \alpha_i L_i$. The conditions on the $\U(1)_X$ charges fulfilling the cancellation of gauge anomalies 
	are reported in the text.}
\end{table}

\subsubsection{Anomaly cancellation} 

The $\U(1)_X$ charges are required to cancel all gauge anomalies. This corresponds to the following five conditions: 
\begin{align}
\label{eq:anom1}
\text{Gravity} \times \U(1)_X :&\quad  2 (X_{\L_L} - X_{\L_R}) + (X_{\E_L} - X_{\E_R}) + (X_{\N_L} - X_{\N_R}) 
- \sum_{\alpha = 1}^N X^{\alpha}_{\nu_R} \nonumber \\
&\quad+ \alpha_{e} + \alpha_{\mu} + \alpha_{\tau} = 0 \, , \\
\label{eq:anom2}
\U(1)^3_X :&\quad 2 (X_{\L_L}^3 - X_{\L_R}^3) + (X_{\E_L}^3 - X_{\E_R}^3) + (X_{\N_L}^3 - X_{\N_R}^3) 
- \sum_{\alpha = 1}^N \(X^{\alpha}_{\nu_R} \)^3 
\nonumber \\
&\quad + \alpha^3_{e} + \alpha^3_{\mu} + \alpha^3_{\tau} 
= 0 \, , \\
\label{eq:anom3}
\SU(2)^2_L \times \U(1)_X :&\quad  \frac{1}{2} (X_{\L_L} - X_{\L_R}) + \frac{1}{2} (3 \alpha_B + \alpha_{e} + \alpha_{\mu} + \alpha_{\tau}) = 0 \, , \\ 
\label{eq:anom4}
\U(1)^2_Y \times \U(1)_X :&\quad  2 (\Y - \frac{1}{2})^2 (X_{\L_L} - X_{\L_R}) 
+ (\Y - 1)^2 (X_{\E_L} - X_{\E_R}) + \Y^2 (X_{\N_L} - X_{\N_R}) \nonumber \\ 
&\quad - \frac{1}{2} (3 \alpha_B + \alpha_{e} + \alpha_{\mu} + \alpha_{\tau}) = 0 \, , \\
\label{eq:anom5}
\U(1)_Y \times \U(1)^2_X :&\quad  2 (\Y - \frac{1}{2}) (X_{\L_L}^2 - X_{\L_R}^2) 
+ (\Y - 1) (X_{\E_L}^2 - X_{\E_R}^2) + \Y (X_{\N_L}^2 - X_{\N_R}^2)  
= 0 \, . 
\end{align}

\subsubsection{Renormalizable operators}

Further constraints on the $\U(1)_X$ charges are obtained 
by the requirement that 
the electroweak anomalons 
pick up their mass from the VEV of $H$.
Hence, the Yukawa Lagrangian involving the electroweak anomalon fields is 
(the discussion of neutrino masses is postponed to \sect{sec:numasses})
\begin{align} 
\label{eq:LgenU1} 
- \mathscr{L}_{Y} &= 
y_{1}\bar{\L}_{L} \E_{R}H+y_{2}\bar{\L}_{R} \E_{L}H+y_{3}\bar{\L}_{L} \N_{R}\tilde{H}+y_{4}\bar{\L}_{R} \N_{L}\tilde{H} 
+ \text{h.c.}
\, ,  
\end{align}
with $\tilde H = i \sigma_2 H^*$. 
The extra conditions on the $\U(1)_X$ charges stemming from \eq{eq:LgenU1} read
\begin{align}
\label{eq:yuk1}
&X_{\E_R} = X_{\L_L} 
\, , \\
\label{eq:yuk2}
&X_{\E_L} = X_{\L_R} 
\, , \\
\label{eq:yuk3}
&X_{\N_R} = X_{\L_L} 
\, , \\
\label{eq:yuk4}
&X_{\N_L} = X_{\L_R} 
\, , 
\end{align}
thus reducing the number of independent 
charges to two. 
By substituting \eqs{eq:yuk1}{eq:yuk4} into \eqs{eq:anom1}{eq:anom5}, 
we obtain the following non-trivial conditions 
\begin{align}
\label{eq:condU1Xn1}
X_{\L_R} - X_{\L_L} &= 3 \alpha_B + \alpha_e + \alpha_{\mu} + \alpha_{\tau} \equiv 3 \alpha_{B+L} \, , \\
\label{eq:condU1Xn2}
\sum_{\alpha = 1}^N X^{\alpha}_{\nu_R} &= \alpha_e + \alpha_{\mu} + \alpha_{\tau} \, , \\
\label{eq:condU1Xn3}
\sum_{\alpha = 1}^N \( X^{\alpha}_{\nu_R} \)^3 &= \alpha^3_e + \alpha^3_{\mu} + \alpha^3_{\tau} \, ,    
\end{align}
where we have introduced the shorthand $\alpha_{B+L}$ defined in \eq{eq:condU1Xn1}. 
Note that the condition of cancellation of electroweak anomalies fixes only the difference 
$X_{\L_R} - X_{\L_R}$, leaving us with one free charge that we choose to be $X_{\L_L}$. 
This redundancy is related to the electroweak anomalon number $\U(1)_{A}$, 
corresponding to a common re-phasing of the electroweak anomalon fields.

Other renormalizable operators, which are allowed by the SM gauge symmetry, may or may not be allowed by 
$\U(1)_X$ invariance. For instance, extra Yukawas of the type\footnote{The case $\S \to \S^*$ is trivially 
obtained by replacing $X_\S \to - X_\S$.}
\beq 
\label{eq:yukVL}
- \Delta\mathscr{L}_{Y} = 
y_\L \bar{\L}_{L} \L_{R} \S^{*} + y_\E \bar{\E}_{L} \E_{R}\S + y_\N \bar{\N}_{L} \N_{R}\S + \text{h.c.} \, , 
\eeq
are only permitted for $X_\S = X_{\L_R} - X_{\L_L} = 3 \alpha_{B+L}$. 
These terms would yield an additional vector-like mass to the anomalons after $\U(1)_X$ symmetry breaking. 
Finally, for specific values 
of $\U(1)_Y$ and $\U(1)_X$ charges, the electroweak anomalons can mix with the SM leptons at the renormalizable level. 
The classification of $d=4$ mixing operators is provided in \Table{tab:anomalonmixing}, 
where we emphasized the phenomenologically relevant case $\Y = 2,-1$ (see \sect{sec:higgs}).   
Note that in the presence of mixing operators the electroweak anomalon number is explicitly broken, 
and hence $X_{\L_L}$ gets fixed in terms of the coefficients of the $X$ generator in \eq{eq:genU1X}. 

\begin{table}[!ht]
	\centering
	\begin{tabular}{|l|l|l|}
	\rowcolor{CGray} 
	\hline
	Mixing operator & $\U(1)_Y$ & $\U(1)_X$ \\ 
	\hline 
	$\bar \ell^{i}_L \E_R H$ & $\Y = 0$ & $X_{\L_L} = \alpha_i$ \\ 
         $\bar \ell^{i}_L \E_R \tilde H$ & $\Y = 1$ & $X_{\L_L} = \alpha_i$ \\ 
	\rowcolor{babyblue}
         $\bar \ell^{i}_L (\E_L)^c H$ & $\Y = 2$ & $X_{\L_L} = - \alpha_i -3\alpha_{B+L}$ \\ 
         $\bar \ell^{i}_L (\E_L)^c \tilde H$ & $\Y = 1$ & $X_{\L_L} = - \alpha_i  -3\alpha_{B+L}$ \\ 
         \rowcolor{babyblue}
         $\bar \ell^{i}_L \N_R H$ & $\Y = -1$ & $X_{\L_L} = \alpha_i $ \\
         $\bar \ell^{i}_L \N_R \tilde H$ & $\Y = 0$ & $X_{\L_L} = \alpha_i $ \\ 
         $\bar \ell^{i}_L (\N_L)^c H$ & $\Y = 1$ & $X_{\L_L} = - \alpha_i  -3\alpha_{B+L}$ \\ 
         $\bar \ell^{i}_L (\N_L)^c \tilde H$ & $\Y = 0$ & $X_{\L_L} = - \alpha_i  -3\alpha_{B+L}$ \\ 
         $\bar \L_L e^i_R H$ & $\Y = 0$ & $X_{\L_L} = \alpha_i $ \\ 
         \rowcolor{babyblue}
         $\bar \L_L e^i_R \tilde H$ & $\Y = -1$ & $X_{\L_L} = \alpha_i$ \\ 
         \rowcolor{babyblue} 
         $\bar \L_R (e^i_R)^c H$ & $\Y = 2$ & $X_{\L_L} = - \alpha_i -3\alpha_{B+L}$ \\ 
         $\bar \L_R (e^i_R)^c \tilde H$ & $\Y = 1$ & $X_{\L_L} = - \alpha_i -3\alpha_{B+L}$ \\
         \hline
         $\bar \L_L \nu^\alpha_R H$ & $\Y = 1$ & $X_{\L_L} = X^\alpha_{\nu_R} $ \\ 
         $\bar \L_L \nu^\alpha_R \tilde H$ & $\Y = 0$ & $X_{\L_L} = X^\alpha_{\nu_R} $ \\ 
         $\bar \L_R (\nu^\alpha_R)^c H$ & $\Y = 1$ & $X_{\L_L} = - X^\alpha_{\nu_R}  -3\alpha_{B+L}$ \\ 
         $\bar \L_R (\nu^\alpha_R)^c \tilde H$ & $\Y = 0$ & $X_{\L_L} = - X^\alpha_{\nu_R} -3\alpha_{B+L}$ \\
         $\bar \L_R \ell^i_L \S $ & $\Y = 0$ & $X_{\L_L} = \alpha_i  + X_\S -3\alpha_{B+L}$ \\
        $\bar \L_L (\ell^i_L)^c \S $ & $\Y = 1$ & $X_{\L_L} = -\alpha_i  + X_\S$ \\ 
	$\bar \E_L e^i_R \S $ & $\Y = 0$ & $X_{\L_L} = \alpha_i + X_\S -3\alpha_{B+L}$ \\ 
	$\bar \E_L \nu^\alpha_R \S $ & $\Y = 1$ & $X_{\L_L} = X_{\nu_R}^\alpha + X_\S -3\alpha_{B+L}$ \\ 
	\rowcolor{babyblue} 
	$\bar \E_R (e^i_R)^c \S $ & $\Y = 2$ & $X_{\L_L} =  -\alpha_i + X_\S$ \\ 
	$\bar \E_R (\nu^\alpha_R)^c \S $ & $\Y = 1$ & $X_{\L_L} = -X_{\nu_R}^\alpha + X_\S$ \\ 
	\rowcolor{babyblue}
	$\bar \N_L e^i_R \S $ & $\Y = -1$ & $X_{\L_L} = \alpha_i  + X_\S -3\alpha_{B+L}$ \\ 
	$\bar \N_L \nu^\alpha_R \S $ & $\Y = 0$ & $X_{\L_L} = X_{\nu_R}^\alpha  + X_\S -3\alpha_{B+L}$ \\ 
	$\bar \N_R (e^i_R)^c \S $ & $\Y = 1$ & $X_{\L_L} = -\alpha_i + X_\S$ \\ 
	$\bar \N_R (\nu^\alpha_R)^c \S $ & $\Y = 0$ & $X_{\L_L} = -X_{\nu_R}^\alpha + X_\S$ \\ 
	\hline
	\end{tabular}	
	\caption{\label{tab:anomalonmixing} 
	Renormalizable operators leading to a mixing between electroweak anomalons and SM leptons
	(first column) and required conditions on $\U(1)_Y$ and 
	$\U(1)_X$ charges (second and third columns). 
	For completeness, we also include mixings via RH neutrinos and/or $\S$, 
	whose $\U(1)_X$ charges depend on the mechanism giving mass to neutrinos (see \sect{sec:numasses}). 
	Mixing operators via $\S^*$ are trivially obtained by flipping the sign of $X_\S$ in the third column. 
	}
\end{table}

\subsubsection{Spectrum}

By adding a proper term in the scalar potential, $\Delta V (H, \S)$, 
the following VEV configurations 
are generated
\beq 
\vev{H} = \frac{1}{\sqrt{2}} 
\begin{pmatrix}
0 \\ v
\end{pmatrix} \, , \qquad 
\vev{\S} = \frac{v_X}{\sqrt{2}} \, , 
\eeq
with $v \simeq 246$ GeV and $v_X$ being the order parameter of $\U(1)_X$ breaking. 
The latter is responsible for the mass of the $\U(1)_X$ gauge boson, $\X^{\mu}$, that is
\beq 
\label{eq:massX}
m_{\X} = X_\S g_X v_X \, ,
\eeq
where $g_X$ is the $\U(1)_X$ gauge coupling entering the covariant derivative, 
i.e.~$D^\mu \S = (\partial^\mu + i g_X X_\S \X^{\mu}) \S$.
The scalar field can be expanded around the vacuum as $\S=\frac{v_{X}}{\sqrt{2}} e^{i\xi/v_X} + \ldots$, 
where $\xi$ is the Goldstone boson associated with the massive state $\X$ and 
we neglected the 
radial mode.
After $\U(1)_{X}$ and 
electroweak symmetry breaking the Yukawa terms in $\mathscr{L}_Y + \Delta \mathscr{L}_Y$ 
(see \eq{eq:LgenU1} and \eq{eq:yukVL})
give mass 
to the electroweak anomalons 
(neglecting for simplicity possible mixings with the SM sector)
\begin{align}
\label{eq:Lbaryonmasses}
- \mathscr{L}_{\rm mass}
&= \bar\Psi^{\E}_{L}  \mathcal{M}_{\E} \, \Psi^{\E}_{R} 
+ \bar\Psi^{\N}_L  \mathcal{M}_{\N} \, \Psi^{\N}_R
+ \text{h.c.}
\, , 
\end{align}
which can be cast into 2-flavour Dirac fermions, 
$\Psi^{\E}_{L,R} = (\E_{\L_{L,R}}, \E_{L,R})$ and 
$\Psi^{\N}_{L,R} = (\N_{\L_{L,R}}, \N_{L,R})$, 
with 
\beq
\mathcal{M}_{\E}=
\begin{pmatrix}
m_{\L} & m_{1} \\
m_{2}^{*} & m_{\E} \\
\end{pmatrix} \, , \qquad 
\mathcal{M}_{\N}=
\begin{pmatrix}
m_{\L} & m_{3} \\
m_{4}^{*} & m_{\N} \\
\end{pmatrix} \, , 
\eeq
and 
\beq 
m_{\L,\,\E,\,\N} =  \frac{y_{\L,\,\E,\,\N}}{\sqrt{2}} v_X \, , \qquad 
m_{1,\,2,\,3,\,4} =  \frac{y_{1,\,2,\,3,\,4}}{\sqrt{2}} v \, .
\eeq 
The mass matrices 
are diagonalized via the bi-unitary transformations 
$\Psi^{\E,\,\N}_{R} \to U_{\E,\,\N} \, \Psi^{\E,\,\N}_{R}$ 
and
$\Psi^{\E,\,\N}_{L} \to V_{\E,\,\N} \, \Psi^{\E,\,\N}_{L}$, 
with the unitary matrices entering non-trivially into the gauge currents in the mass basis. 
In the limit 
$y_{\E} = y_{\N}$, 
$y_{1} = y_{3}$,  
$y_{2} = y_{4}$ (and hence $\mathcal{M}_{\E} = \mathcal{M}_{\N}$), 
the Yukawa Lagrangian 
features a custodial symmetry which helps in taming corrections to 
electroweak precision observables (see \sect{sec:EWPT}). 
In the following, we will stick  
to the custodial limit, 
while for the calculations in \app{app:WZterms} we will consider the more general case.

\subsection{EFT of a light vector and decoupling of WZ terms}
\label{sec:EFTdecWZ}

We are interested in the limit where the electroweak anomalons are heavier the electroweak scale, 
while the vector $\X$ is much lighter than the electroweak scale. 
Parametrically (see \eq{eq:massX}), this can be obtained 
in two ways: 
$i)$ $v_X \gtrsim v$ and $g_X \ll 1$
or 
$ii)$ $v_X \ll v$ and $g_X \lesssim 1$. 
In case $ii)$ or if the $\Delta\mathscr{L}_{Y}$ 
operators in \eq{eq:yukVL} are absent due to charge assignment (i.e.~$X_\S \neq 3 \alpha_{B+L}$), 
this requires $y_{1,2,3,4} \sim \sqrt{4\pi}$ in order for the anomalons to be heavier than 
the electroweak 
scale. 
Upon integrating out the electroweak anomalons at one loop one finds in the EFT 
given by the SM and the light vector $\X$
(also keeping the Goldstone mode $\xi$, see \app{app:WZterms} for details) 
\begin{align}
\label{EFT}
\mathscr{L}_{\rm EFT}^{\U(1)_X} &\supset
g_{X}g'^{2}\frac{C_{BB}}{24\pi^{2}}\epsilon^{\alpha\mu\nu\beta}\X_{\alpha}B_{\mu}\partial_{\beta}B_{\nu}+g_{X}g^{2}\frac{C_{ab}}{24\pi^{2}}\epsilon^{\alpha\mu\nu\beta}\X_{\alpha}W_{\mu}^{a}\partial_{\beta}W_{\nu}^{b} \nonumber \\
&+g_{X}gg'\frac{C_{aB}}{24\pi^{2}}\epsilon^{\alpha\mu\nu\beta}\X_{\alpha}W_{\mu}^{a}\partial_{\beta}B_{\nu}+g_{X}gg'\frac{C_{Ba}}{24\pi^{2}}\epsilon^{\alpha\mu\nu\beta}\X_{\alpha}B_{\mu}\partial_{\beta}W_{\nu}^{a} \, \nonumber\\
&+g_{X}g^{2}\frac{D_{ab}}{48\pi^{2}}\frac{\xi}{m_{\X}}\epsilon^{\alpha\mu\beta\nu}(\partial_{\alpha}W_{\mu}^{a})(\partial_{\beta}W_{\nu}^{b})+g_{X}g'^{2}\frac{D_{BB}}{48\pi^{2}}\frac{\xi}{m_{\X}}\epsilon^{\alpha\mu\beta\nu}(\partial_{\alpha}B_{\mu})(\partial_{\beta}B_{\nu})\, \nonumber\\
&+g_{X}gg'\frac{D_{aB}}{24\pi^{2}}\frac{\xi}{m_{\X}}\epsilon^{\alpha\mu\beta\nu}(\partial_{\alpha}W_{\mu}^{a})(\partial_{\beta}B_{\nu})\, , 
\end{align}
where $a,b = 1,2,3$ and we neglected non-abelian $W$ terms scaling with an extra gauge coupling $g$. 
In general, from the requirement that the 
electromagnetic group remains unbroken, 
one obtains 
\begin{align}
C_{ab}&=\begin{pmatrix}
C_{11} & C_{12} & 0 \\
-C_{12} & C_{11} & 0 \\
0 & 0 & C_{33} 
\end{pmatrix} \,, \quad
C_{aB}=\begin{pmatrix}
0, & 0, & C_{3B} 
\end{pmatrix}\,, \quad 
C_{Ba}=\begin{pmatrix}
0, & 0, & C_{B3} 
\end{pmatrix}\,, \\
D_{ab}&=\begin{pmatrix}
D_{11} & 0 & 0 \\
0 & D_{11} & 0 \\
0 & 0 & D_{33} 
\end{pmatrix} \,, \quad
D_{aB}=\begin{pmatrix}
0, & 0, & D_{3B} 
\end{pmatrix} \,, 
\end{align}
together with the sum-rules 
\begin{align}
&C_{33}+C_{3B}+C_{B3}+C_{BB}=0 \,, \\ 
&D_{33}+2D_{3B}+D_{BB}=0  \,.
\end{align} 
A relatively simple case is given in the limit where 
the masses of the anomalon fields stem completely from the VEV of $\S$, 
yielding 
\begin{align}
C_{11}&=C_{33}=-C_{BB}= 
3\alpha_{B+L} 
\,,\\
C_{B3}&=-C_{3B}=D_{3B}=C_{12}=0 \,,\\
D_{11}&=D_{33}=-D_{BB}=-9 \alpha_{B+L} \, ,
\end{align} 
where the effective coefficients are set by the anomalous trace 
of the SM current 
(see \eqs{simplified}{simplified2}). 
Here, instead, we focus on the more general case where the anomalon masses have both a SM-singlet and an electroweak 
symmetry breaking source. Although we were not able to cast explicit expressions for the EFT coefficients 
into a simple analytical form 
(see \eqs{eq:Cgeneral}{eq:Dgeneral}), 
we will present them here under the simplified 
(but phenomenologically motivated) 
hypothesis 
in which 
the anomalon masses are degenerate, 
that is
\beq
\label{eq:equalmasses}
\mathcal{M}_{\E}^{\dag}\mathcal{M}_{\E}=\mathcal{M}_{\N}^{\dag}\mathcal{M}_{\N}=m_{\Psi}^{2}\begin{pmatrix} 1 & 0 \\ 0 & 1 
\end{pmatrix} \,,
\eeq
with $m_{\Psi}$ denoting the degenerate anomalons mass. \eq{eq:equalmasses} enforces the mass matrices to be
\beq
\label{amuchina}
\mathcal{M}_{\E}=\mathcal{M}_{\N}=\frac{1}{\sqrt{2}}\begin{pmatrix}
y_{\S}v_{X} & iy_{H}v \\
iy_{H}v & y_{\S}v_{X}
\end{pmatrix}=m_{\Psi}\begin{pmatrix}
\cos\theta & i\sin\theta \\
i\sin\theta & \cos\theta
\end{pmatrix} \, ,
\eeq
with $y_{\S,H}$ real parameters,\footnote{The accidental global $\U(1)^{6}$ symmetry corresponding to the re-phasing of the each electroweak anomalon field
is broken by the $\mathscr{L}_{Y} + \Delta\mathscr{L}_{Y}$, leaving an unbroken electroweak anomalon number $\U(1)_A$, 
that is the subgroup corresponding to the common re-phasing of all the anomalon fields. 
Hence, of the seven complex parameters introduced in $\mathscr{L}_{Y} + \Delta\mathscr{L}_{Y}$, $6-1 = 5$ phases 
are unphysical and can be rotated away. One possible choice is to set 
$\text{Arg}(y_{\L})=\text{Arg}(y_{\E})=\text{Arg}(y_{\N})=0$, 
$\text{Arg}(y_{1})=-\text{Arg}(y_{2})$, 
and 
$\text{Arg}(y_{3})=-\text{Arg}(y_{4})$.} 
while 
\beq  
m^2_{\Psi} = \frac{1}{2} \( (y_H v)^2 + (y_S v_X)^2 \) \, ,
\eeq
and 
\beq 
\label{eq:tantheta}
\tan\theta = \frac{y_H v}{y_S v_X} \, . 
\eeq
Specializing the general expressions in \eqs{eq:Cgeneral}{eq:Dgeneral} to the degenerate case above, we find 
\begin{align}
\label{eq:C11}
C_{11}&=C_{33}=-C_{BB}=\frac{3}{4} \alpha_{B+L} (1+3\cos2\theta) \, ,\\
\label{eq:CB3}
C_{B3}&=-C_{3B}=\frac{9}{4} \alpha_{B+L} (1-\cos2\theta),\,\,\,\,\,\,\,\,\,C_{12}=0 \, ,\\
\label{eq:D11}
D_{11}&=D_{33}=-D_{BB}=-\frac{9}{2} \alpha_{B+L} (1+\cos2\theta),\,\,\,\,\,\,\,D_{3B}=0 \, .
\end{align}
The important point to be noted is that 
when the anomalons pick-up a mass from both electroweak preserving and breaking 
sources, the low-energy WZ coefficients acquire a model dependence through the angle $\theta$. 

In order to understand the phenomenological 
implications of this model dependence, we briefly recall here the argument of 
Refs.~\cite{Dror:2017ehi,Dror:2017nsg} regarding the energy-enhanced emission of the longitudinal modes of $\X$ 
stemming from the WZ operators.  
Taking the limit $g_{X}\rightarrow0$ and $m_{\X}\rightarrow0$, 
while keeping fixed the ratio $m_{\X} / g_{X} \propto v_X$,  
the transverse modes of $\X$ decouple,  
while the longitudinal mode is enhanced as $E / m_{\X}$. 
In this regime, the equivalence theorem states that the longitudinally polarized vectors are equivalent 
to the corresponding scalar Goldstone bosons. 
This is readily seen by working in the 
so-called ``Equivalent Gauge'' of Ref.~\cite{Wulzer:2013mza}, 
where the longitudinally polarized state, $|\X_{L}\rangle$, is represented as 
\beq
\langle0|\X_{\mu}(x)|\X_{L}(\vec{p})\rangle=\epsilon_{\mu}^{L}(\vec{p})e^{-ipx} \, , 
\qquad
\langle0|\xi(x)|\X_{L}(\vec{p})\rangle=-ie^{-ipx} \, ,
\eeq
with the 
polarization vector
\beq
\epsilon_{\mu}^{L}(\vec{p})=-\frac{m_{\X}}{E_{\vec{p}}+|\vec{p}|}\left\{1,\frac{\vec{p}}{|\vec{p}|}\right\} \, ,
\eeq
vanishing in the $m_{\X}\rightarrow0$ limit. 
The advantage of this representation is that it makes the equivalence theorem 
explicit, since in the high-energy limit (or equivalently $m_{\X}\rightarrow0$) only the 
Goldstone mode survives.  
Hence, adopting the above prescription, 
only the diagrams with one external $\xi$ contribute to physical processes 
in the $m_{\X}\rightarrow0$ limit. 
For instance, upon integrating out the $W$ boson, the axion-like operator $\xi W^{-}\tilde{W}^{+}$ 
proportional to $D_{11}$
in \eq{EFT} yields the effective interaction 
\beq
\label{EFT FCNC}
g_{\xi d_{i}d_{j}}\bar{d}_{j}\gamma^{\mu}P_{L}d_{i}\,(\partial_{\mu}\xi/m_{\X})\,+\text{h.c.} \, , 
\eeq
in terms of the effective coupling \cite{Izaguirre:2016dfi}
\beq 
\label{eq:gxididj}
g_{\xi d_{i}d_{j}}=-\frac{g_{X}g^{4}}{(4\pi)^{4}}D_{11}\sum_{\alpha=u,c,t}V_{\alpha i}V_{\alpha j}^{*}F(m_{\alpha}^{2}/m_{W}^{2}) \, ,
\eeq
with $D_{11} \propto (1+\cos 2\theta)$ given in \eq{eq:D11}, $V$ denoting the CKM matrix and the loop function
\beq 
F(x)=\frac{x(1+x(\ln x-1))}{(1-x)^{2}} \, .
\eeq
This leads to FCNC processes, such as $K \to \pi \X_L$, 
$B \to K \X_L$, etc,   
whose rate is enhanced as $(E/m_{\X})^{2}$, where $E$ is the decay energy (cf.~the derivative operator in \eq{EFT FCNC}),   
thus implying strong bounds on light 
vector bosons coupled to anomalous currents \cite{Dror:2017ehi,Dror:2017nsg}. 

On the other hand, the above constraints from energy-enhanced $\X_L$ emission disappear for $D_{11} = 0$, 
that is when the $\U(1)_X$ Goldstone decouples from the electroweak anomalons. 
This corresponds to $\theta = \pi/2$, 
which implies that the anomalon masses are entirely due to the Higgs VEV (cf.~\eq{eq:tantheta}). From a top-down 
perspective this condition can be neatly obtained in terms of $\U(1)_X$ charges ($X_\S \neq 3 \alpha_{B+L}$)  
which forbid the operators of $\Delta\mathscr{L}_{Y}$ 
in \eq{eq:yukVL}. 
Alternatively, it can be parametrically obtained  
by taking $v_X \ll v$ or $y_{\S} \approx 0$. 
Note that the latter condition is radiatively stable, since it 
corresponds to an enhanced U(1) global symmetry of the Lagrangian in which LH and RH anomalons fields are rotated with an opposite phase.    

In conclusion,  
we have shown that 
the bounds of Refs.~\cite{Dror:2017ehi,Dror:2017nsg} can be relaxed 
by assuming that the anomalon fields are mostly chiral (namely, their mass mostly stems from the 
Higgs VEV). This possibility, however, leads to non-decoupling signatures in Higgs observables 
and direct searches, to be discussed in \sect{sec:anomalonspheno}. 

\subsection{Neutrino masses}
\label{sec:numasses}

If the $X$ generator has a non-trivial projection on family lepton numbers, $\alpha_{i} \neq 0$ ($i = e,\mu,\tau$), 
then we need RH neutrinos, $\nu_R^{\alpha}$ ($\alpha = 1,\ldots,N$), in order to cancel $\U(1)_X$ and $\U(1)^3_X$ anomalies 
(cf.~conditions in \eqs{eq:condU1Xn2}{eq:condU1Xn3}). 
The simplest solution is to introduce one RH neutrino for each $\alpha_i \neq 0$ and set 
$X^{i}_{\nu_R} = \alpha_i$. 
Another possibility is to have universal charges $X_{\nu_R}^{\alpha} = X_{\nu_R}$, 
so that the anomaly-free conditions are 
\begin{align}
X_{\nu_R} = \( \frac{\alpha^3_e + \alpha^3_{\mu} + \alpha^3_{\tau}}{\alpha_e + \alpha_\mu + \alpha_\tau} \)^{1/2} \, ,  \qquad
N = \( \frac{(\alpha_e + \alpha_\mu + \alpha_\tau)^{3}}{\alpha^3_e + \alpha^3_{\mu} + \alpha^3_{\tau}} \)^{1/2} \, .  
\end{align}
The SM-singlet states $\nu_R^\alpha$ can be used to give mass to light neutrinos via the seesaw mechanism. 
In fact, SM gauge invariance would allow the operators\footnote{We neglect here bare Majorana mass terms, 
since in that case RH neutrinos would not contribute to the cancellation of $\U(1)_X$ and $\U(1)^3_X$ anomalies.  
Instead, possible mixings between RH neutrinos and electroweak anomalons have been classified in \Table{tab:anomalonmixing}.} 
\beq 
\label{eq:LYR}
-\mathscr{L}_{Y}^{\nu_R} = y^{i\beta}_D \bar \ell^i_L \nu_R^\beta \tilde H + \frac{1}{2} y^{\alpha\beta}_{\nu_R} \nu_R^{\alpha} \nu_R^{\beta} \S^* + \text{h.c.} 
\quad \longrightarrow \quad 
m_D^{i\beta} \, \bar \ell^i_L \nu_R^\beta + \frac{1}{2} M_R^{\alpha\beta} \nu_R^{\alpha} \nu_R^{\beta} + \text{h.c.}
\, , 
\eeq
with $m_D = y_D v / \sqrt{2}$ and $M_R = y_R v_X / \sqrt{2}$, leading to light neutrino masses via the seesaw mechanism 
\beq 
m_\nu = m_D M_R^{-1} m_D^T \, .
\eeq
However, in order for the operators in \eq{eq:LYR}  
to be $\U(1)_X$ invariant, the following constraints on $\U(1)_X$ charges need to be satisfied
\begin{align} 
-\alpha_i + X^\beta_{\nu_R} &= 0 \, , \\
X^\alpha_{\nu_R} + X^\beta_{\nu_R} - X_\S &=0 \, .
\end{align}
While the first condition can be easily fulfilled (since it also ensures the cancellation of $\U(1)_X$ and $\U(1)^3_X$ anomalies), 
the second one could imply texture zeros in $M_R$ if some leptonic generators are non-universal $\alpha_i \neq \alpha_j$. 
Consistency with light neutrino data might then require the introduction of extra scalars in order to reproduce realistic 
low-energy textures (see e.g.~\cite{Asai:2019ciz,Altmannshofer:2019xda,Araki:2019rmw}).

\section{Electroweak anomalons phenomenology}
\label{sec:anomalonspheno}

In the previous Section we have seen that mostly chiral electroweak anomalons 
(i.e.~which take their mass mostly from the Higgs VEV) allow to decouple dangerous 
WZ terms, which would otherwise lead to the energy-enhanced longitudinal emission of light vectors coupled to anomalous currents. 
We are hence interested in studying the electroweak phenomenology of the exotic leptons $\L + \N + \E$, 
whose quantum numbers are  
displayed in \Table{tab:genU1fieldcontent}. 
In particular, following the analysis of Refs.~\cite{Bizot:2015zaa,Bonnefoy:2020gyh}, 
we will argue that phenomenology requires 
$\Y \approx 2,-1$.

\subsection{Electroweak precision tests}
\label{sec:EWPT}

The contribution of the electroweak anomalons 
in terms of mass eigenstates 
(cf.~\Table{tab:genU1fieldcontent} and \eq{eq:Lbaryonmasses}) 
to the $S$ and $T$ parameters
is \cite{Bizot:2015zaa}
\begin{align}
S&= \frac{1}{6\pi} 
\[ \(1 - 2 (\Y - \frac{1}{2}) \log\frac{m^2_{\N_1}}{m^2_{\E_1}}\) 
+ \(1 + 2 (\Y - \frac{1}{2}) \log\frac{m^2_{\N_2}}{m^2_{\E_2}}\) 
+ \mathcal{O}\(\frac{m^2_{Z}}{m^2_{\N,\, \E}}\)
\] 
\approx \frac{1}{3\pi} \, , \\
T&= \frac{1}{16\pi c^2_W s^2_W m^2_Z} 
\( m^2_{\N_1} + m^2_{\E_1} 
- 2 \frac{m^2_{\N_1}m^2_{\E_1}}{m^2_{\N_1}-m^2_{\E_1}} \log\frac{m^2_{\N_1}}{m^2_{\E_1}}\) \nonumber \\
&+ \frac{1}{16\pi c^2_W s^2_W m^2_Z} 
\( m^2_{\N_2} + m^2_{\E_2} 
- 2 \frac{m^2_{\N_2}m^2_{\E_2}}{m^2_{\N_2}-m^2_{\E_2}} \log\frac{m^2_{\N_2}}{m^2_{\E_2}}\) 
\approx 0
\, ,
\end{align}
where the approximation in the last steps 
holds in the custodial limit 
$m_{\N_{1,2}} \approx m_{\E_{1,2}}$.
Recent fits for oblique parameters, e.g.~from Gfitter \cite{Gfitter}, yield
\beq 
S= 0.05 \pm 0.11 \, , \qquad 
T= 0.09 \pm 0.13 \, ,  
\eeq
which are easily satisfied in the custodial limit, although a mass splitting might 
play a role to explain the recent $M_W$ anomaly \cite{CDF:2022hxs}.

\subsection{Higgs physics}
\label{sec:higgs}

We now consider the constraints from Higgs couplings measurements. In particular, we assess the impact of the new heavy fermions on the decay rate of the Higgs boson to two photons, or to a photon and a $Z$ boson.
Taking a fermion $\psi$ of mass $m_\psi$ the interaction Lagrangian is given by 
\beq
\cL_\psi^{\rm int}=-\frac{m_\psi}{v}h\overline{\psi}\psi+eQ_\psi\overline{\psi}\gamma^\mu\psi A_\mu+\frac{e}{c_Ws_W}\overline{\psi}\gamma^\mu\left(\frac{T^3_\psi}{2}-Q_\psi s_W^2-\frac{T^3_\psi}{2}\gamma_5\right)\psi Z_\mu \ ,
\label{lagHiggsCouplings}
\eeq
where $h$ is the $125$ GeV Higgs, $A_\mu$ and $Z_\mu$ the photon and $Z$ boson fields, $T^3_\psi$ is the eigenvalue of the third generator of $\SU(2)_L$ when it acts on the left-handed component of $\psi$, so that $T^3_\psi=\pm\frac{1}{2}$ when $\psi_L$ arises from a doublet in the fundamental of $\SU(2)_L$. Its one-loop contributions to the amplitudes $h\to\gamma\gamma$ and $h\to\gamma Z$ are \cite{Djouadi:2005gi}
\beq
\cA^\psi_{\gamma\gamma}\approx\frac{4}{3} Q_\psi^2 \ , \quad \cA^\psi_{Z\gamma}\approx-\frac{1}{3} Q_\psi\frac{T^3_\psi-2Q_\psi s_W^2}{c_W} \ ,
\label{higgsAmplitudeFormulae}
\eeq
where we assumed that $\psi$ is much heavier than the Higgs and the $Z$ boson, which holds for the heavy fermions we consider here. In the SM, these amplitudes are dominated by the loop of the $W$ gauge boson interfering negatively with the loop of the top quark and they amount to $A_{\gamma\gamma}^{\rm SM} \approx -6.5$ and $\mathcal{A}_{\gamma Z}^{\rm SM} 
\approx 5.7$ at leading order. 
In the presence of a single Higgs doublet, the new physics contribution yields $\mathcal{A_{\gamma\gamma}^{\rm NP}} 
\approx \frac{8}{3} (1 - 2\Y +2 \Y^2)$. Writing the modified Higgs width to photons as 
\beq 
R_{\gamma\gamma} = \frac{|\mathcal{A_{\gamma\gamma}^{\rm SM}} + \mathcal{A_{\gamma\gamma}^{\rm NP}|}^2}{\mathcal{|A_{\gamma\gamma}^{\rm SM}|}^2} \, , 
\eeq
a recent ATLAS analysis found $R_{\gamma\gamma} = 1.00 \pm 0.12 $~\cite{Aad:2019mbh}. 
There is in fact the possibility that the new physics contribution interferes negatively with the SM 
amplitude, namely $\mathcal{A_{\gamma\gamma}^{\rm NP}} \approx -2 
\mathcal{A}_{\gamma\gamma}^{\rm SM} \approx 13.0$. 
This is obtained either for $\Y \approx 2$
($1.93 \lesssim \Y \lesssim 2.03$ [2$\sigma$ range]) 
or 
$\Y \approx -1$ ($-1.03 \lesssim \Y \lesssim -0.93$ [2$\sigma$ range]), 
both yielding 
$\mathcal{A_{\gamma\gamma}^{\rm NP}}(\Y = 2) = \mathcal{A_{\gamma\gamma}^{\rm NP}}(\Y = -1) \approx 13.3$. 
A correlated signal in the $\gamma Z$ channel yields $\mathcal{A}_{\gamma Z}^{\rm NP} 
\approx -\frac{2}{3}c_W [1 - (3 - 8 \Y + 8 \Y^2) t^2_{W}]$, leading to a large deviation 
$\mathcal{A}_{\gamma Z}^{\rm NP} (\Y = 2) = \mathcal{A}_{\gamma Z}^{\rm NP} (\Y = -1) \approx 2.33$
in the region 
where the value of $\Y$ is compatible with the di-photon channel. The $\gamma Z$ decay channel of the Higgs has not been observed yet and HL-LHC is expected to measure $\kappa_{\gamma Z}$ within $10\%$ precision~\cite{Cepeda:2019klc}. 
Thus the model with a single Higgs doublet 
predicts a strong departure of $R_{Z\gamma}$ from its SM value, although extended Higgs sectors 
can help to tame modifications of Higgs signals (see e.g.~\cite{Bonnefoy:2020gyh}).

\subsection{Direct searches}
\label{sec:directsearches}

Direct searches at high-energy particle colliders depend on whether the exotic leptons mix with the SM leptons. In fact, 
this is possible only for the values $\Y = 0,\pm 1, 2$ (see \Table{tab:anomalonmixing}), 
including the phenomenologically 
favoured case $\Y = 2,-1$. 
We discuss in turn the two different scenarios 
corresponding to 
$\Y \neq 2,-1$ (stable charged leptons) and $\Y = 2,-1$ (unstable charged leptons).

\subsubsection{Stable charged leptons}

For $\Y \approx 2,-1$ (but $\Y \neq 2,-1$) the exotic leptons do not mix the SM ones and the lightest state of the spectrum 
is electrically charged 
and stable due to 
exotic lepton number. 
Charged relics are cosmologically dangerous and largely excluded. To avoid 
cosmological problems one has to invoke low-scale inflation, such that charged relics are 
either diluted by inflation or never thermally produced. 
On the other hand, stable charged particles yield striking signatures at colliders 
in the form of charged track, anomalous energy loss in calorimeters, longer times of flight, etc. 
Applying the experimental limits of \cite{Khachatryan:2016sfv} at 13 TeV LHC
with the leading-order Drell-Yann cross-sections rescaled for $|Q|=2$
(see also \cite{DiLuzio:2015oha}),
Ref.~\cite{Bonnefoy:2020gyh} obtained $m_{\N,\, \E} \gtrsim 800$ GeV. 
Since $m_{\N,\, \E} = y_{\N,\, \E} \, v/ \sqrt{2}$, direct searches imply Yukawa couplings, 
$y_{\N,\, \E} \approx 4.6$, 
at the boundary 
of perturbative unitarity (see e.g.~\cite{DiLuzio:2016sur,Allwicher:2021rtd}).\footnote{Large higher-order corrections 
(starting at two loops) are expected for Higgs decays and they might slightly change the solutions $\Y \approx 2,-1$.} 

\subsubsection{Unstable charged leptons}

For $\Y = 2,-1$ the electroweak anomalons have electric charge $Q = 2,-1$ ($\N$ components) 
and $Q = 1,-2$ ($\E$ components). The $|Q| = 2$ states can decay into a $W$ and a $|Q| = 1$ fermion, 
while the latter can mix with SM leptons and decay into $Z\ell$ or $h \ell$. 
Signatures of this type were previously studied in Ref.~\cite{Ma:2014zda}, 
which estimated a mass reach at the LHC 14 up to $m_{\N,\, \E} \sim 800$ GeV (depending on the integrated luminosity). 
To our knowledge, however, such an analysis has never been performed by the experimental collaborations. 
Anyhow, the bounds appear to be of the same order of those obtained in the case of stable charged leptons.

\section{Conclusions}
\label{sec:conclusions}

In this work we have revisited the case of light vector bosons coupled to anomalous 
currents which are UV completed by new anomaly-cancelling heavy fermions (anomalons).  
After the latter have been integrated out, 
WZ terms of the type in \eq{EFT} are generated. On the one hand, they take care of anomaly cancellation in the IR and,  
on the other, they source the energy-enhanced emission of longitudinally polarized vectors, $\X$, which 
typically results in very strong bounds on $g_X / m_\X \propto 1/v_X$ 
whenever the decay channels 
$Z \to \gamma \X$, $B \to K \X$, $K \to \pi \X$, etc, are kinematically open 
\cite{Dror:2017ehi,Dror:2017nsg}.  
Here, we have studied the model-dependence of such bounds, considering as a paradigmatic framework the 
gauging of the most general (anomalous) linear combination of SM global symmetries, $\U(1)_X$, with the 
generator $X$ given in \eq{eq:genU1X}.  
To this end, we provided a UV completion including electroweak anomalons $\L + \E + \N$ (cf.~\Table{tab:genU1fieldcontent})
to cancel $\U(1)_X$ anomalies in combination with electroweak gauge factors and RH neutrinos to take care of 
 $\U(1)_X$ anomalies in isolation when the lepton number generators are gauged. 
An extra scalar $\S$ provides the spontaneous breaking of the $\U(1)_X$ factor and gives mass to the vector $\X$. 
Then, we have computed  
the EFT of a light $\X$ when the heavy anomalons are integrated, 
keeping in general both electroweak symmetry breaking and preserving sources for the mass of the anomalons (see \app{app:WZterms} for details). 
This allowed us to conclude (cf.~e.g.~\eq{eq:gxididj}) that the bounds mentioned above  
on light $\X$ can be evaded in the limit where the mass of the electroweak anomalons comes mostly from the 
Higgs VEV. This condition can be neatly imposed in terms of $\U(1)_X$ gauge charges (so that the operators in \eq{eq:LgenU1} are allowed while 
those in \eq{eq:yukVL} are forbidden) or parametrically by decoupling the vector-like masses of the exotic leptons by taking a small yukawa and/or a 
small VEV for $\S$. 
On the other hand, mostly chiral exotic leptons (receiving their mass mostly from the Higgs VEV) are strongly constrained 
due to their non-decoupling nature 
by electroweak-scale phenomenology, in particular Higgs couplings and direct searches. 
We have reviewed in \sect{sec:anomalonspheno} those constraints, 
based on the previous analyses in \cite{Bizot:2015zaa,Bonnefoy:2020gyh}, 
and argued that it is possible to evade $h \to \gamma\gamma$ bounds for $\Y \approx 2,-1$ 
(including the exact cases $\Y = 2,-1$ allowing for mixings between anomalons and SM leptons, cf.~\Table{tab:anomalonmixing}). 
For $\Y \approx 2,-1$, however, the $h \to \gamma Z$ channel differs $\mathcal{O}(1)$ from the SM 
and it will be possible to test this scenario at the HL-LHC. Direct searches, whose signatures depend 
on whether the electroweak anomalons mix or not with the SM leptons, are also very stringent and they practically push the Yukawas 
of the exotic leptons 
to the boundary of perturbativity.   

Since, somewhat surprisingly, mostly chiral exotic leptons are not yet ruled-out, 
it would be interesting to see 
the potential implications of the setup discussed in this paper 
on the physics of light vectors coupled to anomalous currents.  
The most relevant application turns out to be for the case of electro-phobic 
light vectors ($\alpha_e = 0$ in \eq{eq:genU1X}), 
for which the bounds stemming from WZ terms are typically the most important ones. 
These include, for instance, baryonic forces ($\alpha_B \neq 0$) 
which have a rich 
accelerator and collider phenomenology (see e.g.~\cite{Ilten:2018crw,Batell:2021snh}), 
while a case motivated by the $g$-2 anomaly \cite{Muong-2:2021ojo} is that of 
purely muonic forces ($\alpha_\mu \neq 0$) which are in principle 
distinguishable from more standard scenarios based on 
$\U(1)_{\mu - \tau}$ (see e.g.~\cite{Amaral:2021rzw}). 
The general framework discussed in \sect{sec:generalU1} provides a consistent 
UV completion for light vectors coupled to anomalous SM fermionic currents, allowing to free large portions of parameter space from the bounds of 
Ref.~\cite{Dror:2017ehi,Dror:2017nsg}, until the HL-LHC will give the final word on the existence of 
mostly chiral exotic leptons.

\section*{Acknowledgments} 

The work of LDL is partially supported by the European Union's Horizon 2020 research and innovation programme under the Marie Sk\l odowska-Curie grant agreement No 860881 - HIDDEN. 
The work of MN and CT was supported in part by MIUR under contract PRIN 2017L5W2PT

\appendix

\section{Calculation of the Wess-Zumino terms}
\label{app:WZterms}

In this Appendix 
we present the calculation of the effective WZ terms involving gauge and Goldstone bosons that 
arise after integrating 
out heavy fermionic degrees of freedom. In particular, we focus on the 3-point vertices involving the epsilon tensor $\epsilon^{\alpha\beta\mu\nu}$ 
(with $\epsilon^{0123}=1$)
which are related to anomaly cancellation in the EFT when the heavy fermions are integrated out. 

\subsection{Toy model}

We assume a toy model with a set of gauge bosons $G_{\mu}^{A}$ 
related to 
the generators 
$Q^{A}$ 
of the gauge symmetry group $\mathcal{G}$ (that can be in general semi-simple). 
The model contains a fermionic sector, whose fields are labeled as $\psi_{i}$, that acquire a mass term $\mathcal{M}_{ij}$ after a spontaneously symmetry breaking (SSB) mechanism. The $(\frac{1}{2},0)$ and $(0,\frac{1}{2})$ Lorentz components of the $\psi$ field are separately (reducible) representations of 
$\mathcal{G}$
and the generators act on them as
\beq
Q^{A}\psi_{i}=\sum_{j}(Q_{L}^{A})_{ij}\psi_{jL}+\sum_{j}(Q_{R}^{A})_{ij}\psi_{jR} \, ,
\eeq
where $(Q_{L,R}^{A})_{ij}$ are the matrix rapresentation of the gauge multiplets $\psi_{L,R}\equiv P_{L,R}\psi$. 
We restrict ourselves to models with a $\U(1)_{\psi}$ symmetry corresponding to the fermionic number of the 
$\psi$ fields ($\psi_{i}\to e^{i\phi}\psi_{i}$).
The real scalar Higgs fields, responsible for the SSB mechanism, are labeled as $H_{a}=(H_{a})^{*}$ 
and belong to a (reducible) representation of the gauge group $\mathcal{G}$. 
By performing an infinitesimal transformation of angle $\alpha_{A}$ along the $Q^{A}$ generator, the $H_{a}$ fields transform like
\beq
\delta H_{a}=\sum_{b}g_{A}\alpha_{A}(iQ_{H}^{A})_{ab}H_{b} \, ,
\eeq
where $(iQ_{H}^{A})_{ab}$ is a real and antisymmetric matrix.
Hence,  
\beq
\mathcal{L}_{\text{toy}\,\text{model}}\supset\sum_{i}\bar{\psi}_{i}i\slashed{\partial}\psi_{i}-\sum_{a,i,j}H_{a}(\bar{\psi}_{iL}\mathcal{Y}_{ij}^{a}\psi_{jR}+\text{h.c.})-\sum_{A}g_{A}G_{\mu}^{A}J^{\mu A} \, ,
\eeq
with
\beq
J^{\mu A}=\sum_{i,j}\Bigl[\bar{\psi}_{iL}\gamma^{\mu}(Q_{L}^{A})_{ij}\psi_{jL}+\bar{\psi}_{iR}\gamma^{\mu}(Q_{R}^{A})_{ij}\psi_{jR}\Bigr] \, .
\eeq
The Yukawa couplings must preserve gauge invariance and hence they satisfy
\beq
\sum_{k}\mathcal{Y}_{ik}^{a}(Q_{R}^{A})_{kj}-\sum_{k}(Q_{L}^{A})_{ik}\mathcal{Y}_{kj}^{a}+\sum_{b}\mathcal{Y}_{ij}^{b}(Q_{H}^{A})_{ba}=0 \, .
\eeq
The Higgs fields acquire the VEVs $\langle H_{a}\rangle=v_{a}$ which break the gauge group, leaving 
an unbroken subgroup $\mathcal{G}_{0}$. Then, the mass matrix of the $\psi$ fields is given by
\beq
\mathcal{M}_{ij}=\sum_{a}\mathcal{Y}_{ij}^{a}v_{a} \, ,
\eeq
leading to
\beq
\begin{split}
\mathcal{L}_{\text{toy}\,\text{model}}\supset&\sum_{i}\bar{\psi}_{i}i\slashed{\partial}\psi_{i}-\sum_{i,j}(\bar{\psi}_{iL}\mathcal{M}_{ij}\psi_{jR}+\text{h.c.})\\
&-\sum_{a,i,j}\tilde{H}_{a}(\bar{\psi}_{iL}\mathcal{Y}_{ij}^{a}\psi_{jR}+\text{h.c.})-\sum_{A}g_{A}G_{\mu}^{A}J^{\mu A} \, ,
\end{split}
\eeq
where $\tilde{H}_{a}=H_{a}\!-\!v_{a}$ 
are the Higgs fluctuations around the vacuum.

In order to go in the mass basis, 
the mass matrix $\mathcal{M}$ is diagonalized via the bi-unitary transformations 
$\psi_{R} \to U_{R} \, \psi_{R}$ 
and
$\psi_{L} \to U_{L} \, \psi_{L}$, 
which by construction satisfy $U_{L}^{\dag}\mathcal{M}U_{R}=
\text{diag}(m_{1},m_{2},...)$. This yields
\beq
\mathcal{L}_{\text{toy}\,\text{model}}\supset\sum_{i}\bar{\psi}_{i}(i\slashed{\partial}-m_{i})\psi_{i}-\sum_{A}g_{A}G_{\mu}^{A}J_{U}^{\mu A}-\sum_{a,i,j}\tilde{H}_{a}\bar{\psi}_{i}(\hat{\mathcal{Y}}_{R}^{a}P_{R}+\hat{\mathcal{Y}}_{L}^{a}P_{L})_{ij}\psi_{j} \, ,
\eeq
where $\hat{\mathcal{Y}}_{R}^{a}=U_{L}^{\dag}\mathcal{Y}^{a}U_{R}=(\hat{\mathcal{Y}}_{L}^{a})^{\dag}$, 
while the gauge currents in the mass basis are equal to
\beq
J_{U}^{\mu A}=\sum_{i,j}\Bigl[\bar{\psi}_{iL}\gamma^{\mu}(U_{L}^{\dag}Q_{L}^{A}U_{L})_{ij}\psi_{jL}+\bar{\psi}_{iR}\gamma^{\mu}(U_{R}^{\dag}Q_{R}^{A}U_{R})_{ij}\psi_{jR}\Bigr] \, .
\eeq
After integrating out the heavy fermion fields, we get EFT operators of the type 
\beq
\begin{split}
\label{eq:Ltoyeff}
\mathcal{L}_{\text{toy}\,\text{model}}\supset
&\sum_{A,B,C}\frac{g_{A}g_{B}g_{C}}{48\pi^{2}}C^{ABC}\epsilon^{\alpha\mu\nu\beta}G_{\alpha}^{A}G_{\mu}^{B}\partial_{\beta}G_{\nu}^{C}\\
-&\sum_{a,B,C}\frac{g_{B}g_{C}}{48\pi^{2}}D^{a BC}\epsilon^{\mu\nu\alpha\beta}\tilde{H}_{a}\partial_{\alpha}G_{\mu}^{B}\partial_{\beta}G_{\nu}^{C}\, ,
\end{split}
\eeq
in terms of the EFT coefficients 
$C^{ABC}=-C^{BAC}$ and $D^{a BC}=D^{a CB}$ 
that we want to compute. 
Moreover, integrating by parts the term on the first line of \eq{eq:Ltoyeff}, one also obtains 
\beq
C^{ABC}+C^{CAB}+C^{BCA}=0 \, .
\eeq

\subsection{$\gamma_5$ in dimensional regularization}

Dimensional regularization allows to regularize the divergences arising from loop calculations in 4 dimensions, 
while explicitly preserving Lorentz covariance and gauge invariance.
In the $d$-dimensional spacetime, the mass dimensions of the quantum fields are equal to
\beq
[\psi]=\frac{d-1}{2}\, , \qquad [H]=[G_{\mu}]=\frac{d-2}{2} \, .
\eeq
Hence, in order to keep the gauge couplings dimensionless, one introduces the renormalization scale $\mu$ by the substitution
\beq
g\rightarrow \mu^{\frac{4-d}{2}}\,g
\eeq
in the Lagrangian.
The use of dimensional regularization poses some potential problems in calculations where the $\gamma^{5}$ matrix is involved. In fact, $\gamma^{5}$ (or equivalently the antisymmetric tensor $\epsilon^{\alpha\beta\mu\nu}$) is a quantity whose definition is strictly connected to the fact that space-time is four dimensional, and a definition in $d$ dimensions requires special care. Here, we adopt the Breitenlohner-Maison-'t Hooft-Veltman (BMHV) scheme, 
which is able to reproduce the chiral anomaly (see \cite{Belusca-Maito:2020ala} for a recent review).  

We decompose all matrices into a four-dimensional (denoted by bars) and an extra-dimensional (also called ``evanescent'', denoted by hats) component:
\beq
\gamma^{\mu}=\bar{\gamma}^{\mu}+\hat\gamma^{\mu}\,,
\eeq
where $\bar\gamma^{\mu}$ is non-zero only when $\mu$ takes the ordinary values $0,1,2,3$ and $\hat\gamma^{\mu}$ vanishes 
for $\mu = 0,1,2,3$. 
Correspondingly, the matrix tensor $g_{\mu\nu}$ has a four-dimensional and an extra-dimensional part,
\beq
g_{\mu\nu}=\bar g_{\mu\nu} + \hat g_{\mu\nu}\,, 
\eeq
while mixed components vanish. The gamma matrices satisfy
\beq
\{\bar\gamma^{\mu},\bar\gamma^{\nu}\}=2\bar{g}^{\mu\nu}\, ,\qquad 
\{\hat\gamma^{\mu},\hat\gamma^{\nu}\}=2\hat{g}^{\mu\nu}\, ,\qquad 
\{\bar\gamma^{\mu},\hat\gamma^{\nu}\}=0 \, .
\eeq
Then, we simply define $\gamma^{5}$ as in four dimensions, that is
\beq
\label{gamma5}
\gamma^{5}=i\bar\gamma^{0} \bar\gamma^{1} \bar\gamma^{2} \bar\gamma^{3} \, .
\eeq
It is easy to check that the definition in \eq{gamma5} implies
\begin{gather}
\{\gamma^{5},\bar\gamma^{\mu}\}=0\, , \qquad 
[\gamma^{5},\hat\gamma^{\mu}]=0 \, , \\
\text{Tr}\,\gamma^{5}\gamma^{\alpha} \gamma^{\beta} \gamma^{\mu} \gamma^{\nu}=\text{Tr}\,\gamma^{5}\bar\gamma^{\alpha} \bar\gamma^{\beta} \bar\gamma^{\mu} \bar\gamma^{\nu}=4i\epsilon^{\alpha\beta\mu\nu} \, ,
\end{gather}
which is the correct four-dimensional result.

In a general chiral gauge theory, the fermion fields are introduced as Weyl spinors whose formalism is intrinsically tied to 4-dimensional space. In $d$ dimensions, we replace the Weyl spinors by projections of Dirac spinors, which can be generalized to arbitrary dimensions. The right and left projections are $P_{R,L} = \frac{1}{2} (1 \pm \gamma_5)$, 
as in the 4-dimensional space. Then, there are three possible inequivalent choices for the $d$-dimensional extension of the right-handed chiral current $\bar\psi_{iR}\gamma^{\mu}\psi_{jR}$ coupled to gauge bosons, which are
\beq
\bar\psi_{i}P_{L}\gamma^{\mu}\psi_{j}\, , \qquad 
\bar\psi_{i}\gamma^{\mu}P_{R}\psi_{j}\, , \qquad 
\bar\psi_{i}P_{L}\gamma^{\mu}P_{R}\psi_{j} \, .
\eeq
They are different because $P_{L}\gamma^{\mu}\neq\gamma^{\mu}P_{R}$ in $d$ dimensions. Each of these does lead to valid $d$-dimensional extensions of the model that are perfectly renormalizable using dimensional regularization and the BMHV scheme. However, the intermediate calculations and the final $d$-dimensional results will differ, depending on the choice for this interaction term. Our choice for this work is to use the third option, that is
\beq
\bar\psi_{i}P_{L}\gamma^{\mu}P_{R}\psi_{j}=\bar\psi_{iR}\bar\gamma^{\mu}\psi_{jR} \, ,
\eeq
is the most symmetric one, and leads to the simplest expressions. Similar considerations hold for the left-handed chiral current $\bar\psi_{iL}\gamma^{\mu}\psi_{jL}$.
A different choice has to be taken instead for the kinetic terms $\bar\psi_{iR}i\slashed{\partial}\psi_{iR}$ and $\bar\psi_{iL}i\slashed{\partial}\psi_{iL}$. Indeed, in order to properly regularize the theory, we need to consider the full Dirac fermion kinetic term $\bar\psi_{i}i\slashed{\partial}\psi_{i}$, including the evanescent terms. 

Once the regulated amplitude is well-defined, we can perform all the necessary subtractions of the divergences of its sub-diagrams and the resulting finite expression is interpreted in the physical 4-dimensional space by setting all quantities to their 4-dimensional values, i.e.~first taking the $d\to4$ limit and then, setting all remaining evanescent terms to zero.

\subsection{1-loop matching}

\begin{figure}[ht]
\centering
\includegraphics[width=13cm]{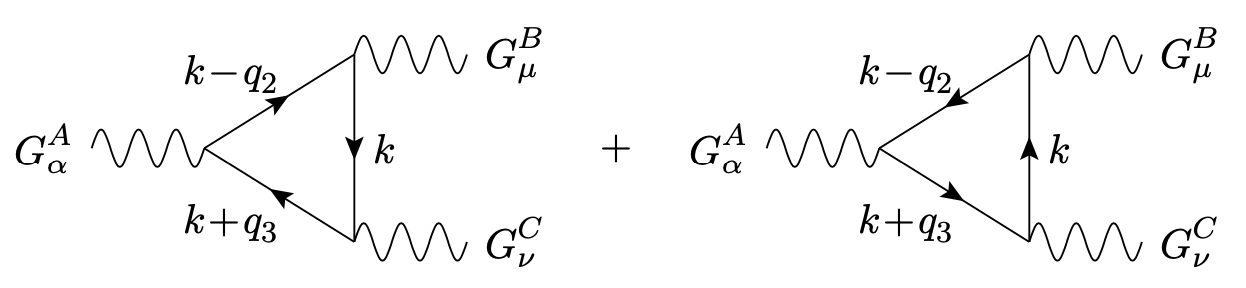} \vspace{0.5cm} \\
\includegraphics[width=13cm]{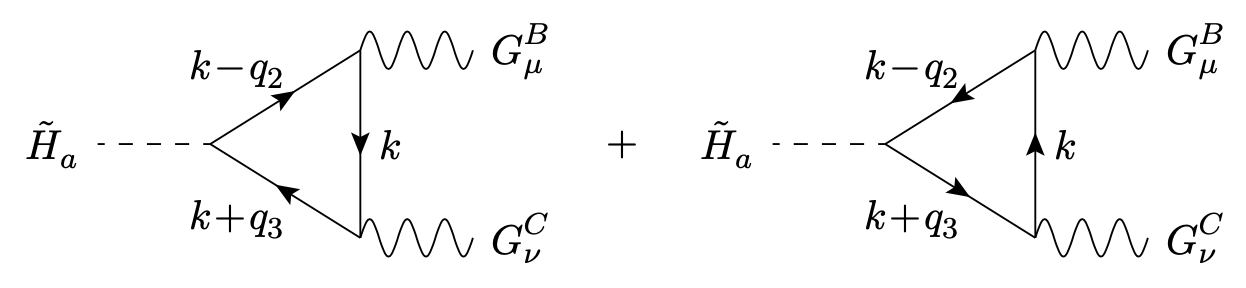}
\caption{Feynman diagrams relative to the 3-point functions in \eqs{eq:MABC}{eq:MaBC}. }
\label{fig:dia3gauge}       
\end{figure}

The epsilon tensor structure occurs in the 3-point functions $\Gamma_{ABC}^{\alpha\mu\nu}(x,y,z)$ and $\Gamma_{a BC}^{\mu\nu}(x,y,z)$ at 1-loop through fermionic triangle diagrams (see \fig{fig:dia3gauge}). The amplitudes in momentum space are defined via  
\beq
\label{eq:MABC}
\int\!\text{d}^{4}\!x\,\text{d}^{4}\!y\,\text{d}^{4}\!z\,e^{i(xq_{1}+yq_{2}+zq_{3})}\,\Gamma_{ABC}^{\alpha\mu\nu}(x,y,z)|_{\text{1-loop}}=(2\pi)^{4}\delta^{(4)}\!(q_{1}+q_{2}+q_{3})\,\mu^{\frac{4-d}{2}}\,iM_{ABC}^{\alpha\mu\nu}(q_{1},q_{2},q_{3}) \, ,
\eeq
and
\beq
\label{eq:MaBC}
\int\!\text{d}^{4}\!x\,\text{d}^{4}\!y\,\text{d}^{4}\!z\,e^{i(xq_{1}+yq_{2}+zq_{3})}\,\Gamma_{a BC}^{\mu\nu}(x,y,z)|_{\text{1-loop}}=(2\pi)^{4}\delta^{(4)}\!(q_{1}+q_{2}+q_{3})\,\mu^{\frac{4-d}{2}}\,iM_{a BC}^{\mu\nu}(q_{1},q_{2},q_{3}) \, , 
\eeq
which yield
\beq
\begin{split}
M_{ABC}^{\alpha\mu\nu}=\sum_{\substack{i,j,k\\ \chi_{1},\chi_{2},\chi_{3}}}\!&g_{A}g_{B}g_{C}(U_{\chi_{1}}^{\dag}Q_{\chi_{1}}^{A}U_{\chi_{1}})_{jk}(U_{\chi_{2}}^{\dag}Q_{\chi_{2}}^{B}U_{\chi_{2}})_{ki}(U_{\chi_{3}}^{\dag}Q_{\chi_{3}}^{C}U_{\chi_{3}})_{ij} \\
&\times i\mu^{4-d}\!\int\!\frac{\text{d}^{d}\!k}{(2\pi)^{d}}\,\frac{\text{Tr}_{\text{D}}[\bar\gamma^{\mu}\!P_{\chi_{2}}\!(\slashed{k}+m_{i})\bar\gamma^{\nu}\!P_{\chi_{3}}\!(\slashed{k}+\slashed{q}_{3}+m_{j})\bar\gamma^{\alpha}\!P_{\chi_{1}}\!(\slashed{k}-\slashed{q}_{2}+m_{k})]}{[k^{2}-m_{i}^{2}]\,[(k+q_{3})^{2}-m_{j}^{2}]\,[(k-q_{2})^{2}-m_{k}^{2}]}\\
+\sum_{\substack{i,j,k\\ \chi_{1},\chi_{2},\chi_{3}}}\!&g_{A}g_{C}g_{B}(U_{\chi_{1}}^{\dag}Q_{\chi_{1}}^{A}U_{\chi_{1}})_{kj}(U_{\chi_{3}}^{\dag}Q_{\chi_{3}}^{C}U_{\chi_{3}})_{ji}(U_{\chi_{2}}^{\dag}Q_{\chi_{2}}^{B}U_{\chi_{2}})_{ik} \\
&\times i\mu^{4-d}\!\int\!\frac{\text{d}^{d}\!k}{(2\pi)^{d}}\,\frac{\text{Tr}_{\text{D}}[\bar\gamma^{\nu}\!P_{\chi_{3}}\!(\slashed{k}+m_{i})\bar\gamma^{\mu}\!P_{\chi_{2}}\!(\slashed{k}+\slashed{q}_{2}+m_{k})\bar\gamma^{\alpha}\!P_{\chi_{1}}\!(\slashed{k}-\slashed{q}_{3}+m_{j})]}{[k^{2}-m_{i}^{2}]\,[(k+q_{2})^{2}-m_{k}^{2}]\,[(k-q_{3})^{2}-m_{j}^{2}]} \, ,
\end{split}
\eeq
and
\beq
\begin{split}
M_{a BC}^{\mu\nu}=\sum_{\substack{i,j,k\\ \chi_{1},\chi_{2},\chi_{3}}}\!&g_{B}g_{C}(\hat{\mathcal{Y}}_{\chi_{1}}^{a})_{jk}(U_{\chi_{2}}^{\dag}Q_{\chi_{2}}^{B}U_{\chi_{2}})_{ki}(U_{\chi_{3}}^{\dag}Q_{\chi_{3}}^{C}U_{\chi_{3}})_{ij}\\
&\times i\mu^{4-d}\!\int\!\frac{\text{d}^{d}\!k}{(2\pi)^{d}}\,\frac{\text{Tr}_{\text{D}}[\bar\gamma^{\mu}\!P_{\chi_{2}}\!(\slashed{k}+m_{i})\bar\gamma^{\nu}\!P_{\chi_{3}}\!(\slashed{k}+\slashed{q}_{3}+m_{j})P_{\chi_{1}}\!(\slashed{k}-\slashed{q}_{2}+m_{k})]}{[k^{2}-m_{i}^{2}]\,[(k+q_{3})^{2}-m_{j}^{2}]\,[(k-q_{2})^{2}-m_{k}^{2}]}\\
+\sum_{\substack{i,j,k\\ \chi_{1},\chi_{2},\chi_{3}}}\!&g_{B}g_{C}(\hat{\mathcal{Y}}_{\chi_{1}}^{a})_{kj}(U_{\chi_{3}}^{\dag}Q_{\chi_{3}}^{C}U_{\chi_{3}})_{ji}(U_{\chi_{2}}^{\dag}Q_{\chi_{2}}^{B}U_{\chi_{2}})_{ik}\\
&\times i\mu^{4-d}\!\int\!\frac{\text{d}^{d}\!k}{(2\pi)^{d}}\,\frac{\text{Tr}_{\text{D}}[\bar\gamma^{\nu}\!P_{\chi_{3}}\!(\slashed{k}+m_{i})\bar\gamma^{\mu}\!P_{\chi_{2}}\!(\slashed{k}+\slashed{q}_{2}+m_{k})P_{\chi_{1}}\!(\slashed{k}-\slashed{q}_{3}+m_{j})]}{[k^{2}-m_{i}^{2}]\,[(k+q_{2})^{2}-m_{k}^{2}]\,[(k-q_{3})^{2}-m_{j}^{2}]}\,.
\end{split}
\eeq
Since we have regularized the theory, the loop integrals over momentum $k$ are convergent and can be evaluated with the usual well-known techniques.
Next, we perform the traces over the Dirac indices to extract the terms involving the epsilon tensor structure we are interested on. One finds that such terms are finite, i.e.~they do not contain 
$1/(d\!-\!4)$
 poles, and are independent from the renormalization scale $\mu$. 
 Hence, we can send $d\rightarrow4$ and set the evanescent components to zero.

In order to obtain the EFT coefficients in \eq{eq:Ltoyeff}, we have to match the expressions that we have calculated above 
to the EFT matrix elements in the limit of heavy fermion masses, i.e.
\beq
\lim_{\substack{m_{i,j,k}^{2}\gg \\ q_{2}^{2},q_{3}^{2},q_{2}\!\cdot\!q_{3}}}M_{ABC}^{\alpha\mu\nu}|_{\epsilon-tensor}=\frac{g_{A}g_{B}g_{C}}{24\pi^{2}}\epsilon^{\alpha\mu\nu\beta}(C^{ABC}iq_{3}+C^{CAB}iq_{2}+C^{BCA}iq_{1})_{\beta} \, ,
\eeq
and
\beq
\lim_{\substack{m_{i,j,k}^{2}\gg \\ q_{2}^{2},q_{3}^{2},q_{2}\!\cdot\!q_{3}}}M_{a BC}^{\mu\nu}|_{\epsilon-tensor}=\frac{g_{B}g_{C}}{24\pi^{2}}D^{a BC}\epsilon^{\mu\nu\alpha\beta}q_{2\alpha}q_{3\beta}\,.
\eeq
Thus we get
\beq
\label{eq:Cgeneral}
\begin{split}
C^{ABC}=\int_{0}^{+\infty}\!\!\!\text{d}s\!\int_{0}^{1}\!\text{d}x\!&\int_{0}^{1}\!\text{d}y\!\int_{0}^{1}\!\text{d}z\,2\,\delta(1\!-\!x\!-\!y\!-\!z)\times\\
\times\,\text{Re}\Biggl\{ 3y\,&\text{Tr}\!\left[e^{-sy\mathcal{M}^{\dag}\mathcal{M}}Q_{R}^{A}\mathcal{M}^{\dag}e^{-sz\mathcal{M}\mathcal{M}^{\dag}}Q_{L}^{B}e^{-sx\mathcal{M}\mathcal{M}^{\dag}}Q_{L}^{C}\mathcal{M}\right]\\
-3y\,&\text{Tr}\!\left[e^{-sy\mathcal{M}^{\dag}\mathcal{M}}Q_{R}^{B}\mathcal{M}^{\dag}e^{-sz\mathcal{M}\mathcal{M}^{\dag}}Q_{L}^{A}e^{-sx\mathcal{M}\mathcal{M}^{\dag}}Q_{L}^{C}\mathcal{M}\right]\\
+3y\,&\text{Tr}\!\left[e^{-sy\mathcal{M}\mathcal{M}^{\dag}}Q_{L}^{B}\mathcal{M}e^{-sz\mathcal{M}^{\dag}\mathcal{M}}Q_{R}^{A}e^{-sx\mathcal{M}^{\dag}\mathcal{M}}Q_{R}^{C}\mathcal{M}^{\dag}\right]\\
-3y\,&\text{Tr}\!\left[e^{-sy\mathcal{M}\mathcal{M}^{\dag}}Q_{L}^{A}\mathcal{M}e^{-sz\mathcal{M}^{\dag}\mathcal{M}}Q_{R}^{B}e^{-sx\mathcal{M}^{\dag}\mathcal{M}}Q_{R}^{C}\mathcal{M}^{\dag}\right]\\
+y\,&\text{Tr}\!\left[e^{-sy\mathcal{M}^{\dag}\mathcal{M}}\mathcal{M}^{\dag}\mathcal{M}Q_{R}^{A}e^{-sz\mathcal{M}^{\dag}\mathcal{M}}Q_{R}^{B}e^{-sx\mathcal{M}^{\dag}\mathcal{M}}Q_{R}^{C}\right]\\
-x\,&\text{Tr}\!\left[e^{-sy\mathcal{M}^{\dag}\mathcal{M}}Q_{R}^{A}e^{-sz\mathcal{M}^{\dag}\mathcal{M}}Q_{R}^{B}e^{-sx\mathcal{M}^{\dag}\mathcal{M}}\mathcal{M}^{\dag}\mathcal{M}Q_{R}^{C}\right]\\
+x\,&\text{Tr}\!\left[e^{-sy\mathcal{M}\mathcal{M}^{\dag}}Q_{L}^{A}e^{-sz\mathcal{M}\mathcal{M}^{\dag}}Q_{L}^{B}e^{-sx\mathcal{M}\mathcal{M}^{\dag}}\mathcal{M}\mathcal{M}^{\dag}Q_{L}^{C}\right]\\
-y\,&\text{Tr}\!\left[e^{-sy\mathcal{M}\mathcal{M}^{\dag}}\mathcal{M}\mathcal{M}^{\dag}Q_{L}^{A}e^{-sz\mathcal{M}\mathcal{M}^{\dag}}Q_{L}^{B}e^{-sx\mathcal{M}\mathcal{M}^{\dag}}Q_{L}^{C}\right]\Biggr\} \, ,
\end{split}
\eeq
and
\beq
\label{eq:Dgeneral}
\begin{split}
D^{a BC}=\int_{0}^{+\infty}\!\!\!\!\!\!\text{d}s\!\int_{0}^{1}\!\text{d}x\!\int_{0}^{1}\!&\text{d}y\!\int_{0}^{1}\!\text{d}z\,6\,\delta(1\!-\!x\!-\!y\!-\!z)\times\\
\times\,\text{Im}\Biggl\{x\,&\text{Tr}\!\left[e^{-sz\mathcal{M}^{\dag}\mathcal{M}}Q_{R}^{B}\mathcal{M}^{\dag}e^{-sx\mathcal{M}\mathcal{M}^{\dag}}Q_{L}^{C}e^{-sy\mathcal{M}\mathcal{M}^{\dag}}\mathcal{Y}^{a}\right]\\
+x\,&\text{Tr}\!\left[e^{-sz\mathcal{M}^{\dag}\mathcal{M}}Q_{R}^{C}\mathcal{M}^{\dag}e^{-sx\mathcal{M}\mathcal{M}^{\dag}}Q_{L}^{B}e^{-sy\mathcal{M}\mathcal{M}^{\dag}}\mathcal{Y}^{a}\right]\\
+y\,&\text{Tr}\!\left[e^{-sy\mathcal{M}^{\dag}\mathcal{M}}\mathcal{M}^{\dag}Q_{L}^{C}e^{-sx\mathcal{M}\mathcal{M}^{\dag}}Q_{L}^{B}e^{-sz\mathcal{M}\mathcal{M}^{\dag}}\mathcal{Y}^{a}\right]\\
+y\,&\text{Tr}\!\left[e^{-sy\mathcal{M}^{\dag}\mathcal{M}}\mathcal{M}^{\dag}Q_{L}^{B}e^{-sx\mathcal{M}\mathcal{M}^{\dag}}Q_{L}^{C}e^{-sz\mathcal{M}\mathcal{M}^{\dag}}\mathcal{Y}^{a}\right]\\
+y\,&\text{Tr}\!\left[e^{-sz\mathcal{M}^{\dag}\mathcal{M}}Q_{R}^{B}e^{-sx\mathcal{M}^{\dag}\mathcal{M}}Q_{R}^{C}\mathcal{M}^{\dag}e^{-sy\mathcal{M}\mathcal{M}^{\dag}}\mathcal{Y}^{a}\right]\\
+y\,&\text{Tr}\!\left[e^{-sz\mathcal{M}^{\dag}\mathcal{M}}Q_{R}^{C}e^{-sx\mathcal{M}^{\dag}\mathcal{M}}Q_{R}^{B}\mathcal{M}^{\dag}e^{-sy\mathcal{M}\mathcal{M}^{\dag}}\mathcal{Y}^{a}\right]\Biggr\}\,.
\end{split}
\eeq

\subsection{Reproducing the chiral anomaly in the EFT}

A consistent gauge theory must be anomaly free and hence the chiral anomaly needs to cancel when we sum over all the fermion fields of the theory. If we integrate out a heavy fermionic sector of the complete UV model, the corresponding chiral anomaly is reproduced in the EFT action $\S_{eff}$ thanks to the 
WZ effective operators in \eq{eq:Ltoyeff}. 
To show this, we make an infinitesimal transformation of angle $\alpha_{A}$ along the $Q^{A}$ generator. The gauge fields $G_{\mu}^{B}$ and the Higgs fields $\tilde{H}_{a}$ transform like
\begin{gather}
\delta \tilde{H}_{a}=\sum_{b}\alpha_{A}(iQ_{H}^{A})_{ab}v_{b}\,+\,\text{linear\,terms}\, ,\\
\delta G_{\mu}^{B}=-\delta_{AB}\,(\partial_{\mu}\alpha_{B})/g_{B}\,+\,\text{linear\,terms}\, ,
\end{gather}
which, from the variation of the effective Lagrangian in \eq{eq:Ltoyeff}, yields 
\beq
\begin{split}
\delta
\S_{eff}
=&\sum_{BC}\frac{g_{B}g_{C}}{48\pi^{2}}\left[C^{ABC}+C^{ACB}+D^{aBC}(iQ_{H}^{A})_{ab}v_{b}\right]\int\!\text{d}^{4}x\,\alpha_{A}\partial_{\alpha}G_{\mu}^{B}\partial_{\beta}G_{\nu}^{C}\epsilon^{\mu\nu\alpha\beta}\\
=&\sum_{BC}\frac{g_{B}g_{C}}{48\pi^{2}}\left[\text{Tr}\,Q_{R}^{A}\{Q_{R}^{B},Q_{R}^{C}\}-\text{Tr}\,Q_{L}^{A}\{Q_{L}^{B},Q_{L}^{C}\}\right]\int\!\text{d}^{4}x\,\alpha_{A}\partial_{\alpha}G_{\mu}^{B}\partial_{\beta}G_{\nu}^{C}\epsilon^{\alpha\mu\beta\nu}\, .
\end{split}
\eeq

\subsection{WZ terms for massive vector bosons and Goldstone bosons}

The VEVs of the Higgs fields contribute to the mass matrix
$\mathcal{M}_{gauge}^{2}$ for the gauge bosons $G_{\mu}^{A}$ with elements
\beq
(\mathcal{M}_{gauge}^{2})_{AB}=\sum_{a,b,c}g_{A}(iQ_{H}^{A})_{ca}v_{a}g_{B}(iQ_{H}^{B})_{cb}v_{b} \,.
\eeq
Since the matrix is real and symmetric, we can diagonalize it through an orthogonal matrix $O_{AB}$ such that
\beq
\sum_{A,B}\,O_{DB}O_{CA}(\mathcal{M}_{gauge}^{2})_{AB}=m_{C}^{2}\delta_{CD} \,.
\eeq
The massive eigenstates are then defined by
\beq
Z_{\mu}^{A}=\sum_{B}\,O_{AB}\,G_{\mu}^{B} \, ,
\eeq
with corresponding symmetry generator
\beq
\tilde{g}_{A}T^{A}=\sum_{B}\,O_{AB}\,g_{B}Q^{B}\,.
\eeq
There are two scenarios for each generator $T^{A}$:
\begin{itemize}
\item $(iT_{H}^{A})_{ab}v_{b}=0$, if $T^{A}$ belongs to the unbroken subgroup $\mathcal{G}_{0}$, such that the corresponding vector boson $Z_{\mu}^{A}$ is then massless, i.e.~$m_{A}=0$;
\item $(iT_{H}^{A})_{ab}v_{b}\neq0$, if $T^{A}$ is spontaneously broken by the Higgs VEVs. The corresponding Nambu-Goldstone boson $\eta^{A}$ is given by
\beq
\eta^{A}=\sum_{a}\,t_{a}^{A}\tilde{H}_{a}\,,
\eeq
where $t_{a}^{A}=\tilde{g}_{A}(iT_{H}^{A})_{ab}v_{b}/m_{A}$ are a (incomplete) set of orthogonal vectors in the Higgs space, i.e.~$\sum_{a}t_{a}^{A}t_{a}^{B}=\delta_{AB}$. The Goldstone field is then eaten by the vector boson $Z_{\mu}^{A}$ which acquires a mass $m_{A}$.
\end{itemize}
The $\tilde{H}_{a}$ contains the Goldstone modes along the $t_{a}^{A}$ directions while the remaining modes, orthogonal to the Goldstones, are all physical, i.e.
\beq
\tilde{H}_{a}=\sum_{\text{NG\,modes}}t_{a}^{A}\eta^{A}+\dots\,.
\eeq
The interaction terms between the $\psi$ fields and the Goldstone bosons are given by
\beq
\sum_{a}\mathcal{Y}_{ij}^{a}t_{a}^{A}=\frac{i\tilde{g}_{A}}{m_{A}}\sum_{k}\left[(T_{L}^{A})_{ik}\mathcal{M}_{kj}-\mathcal{M}_{ik}(T_{R}^{A})_{kj}\right] \, ,
\eeq
because of the gauge invariance of the Yukawa couplings.
Upon an infinitesimal transformation of angle $\alpha_A$ along 
a broken $T^{A}$ generator, the Goldstone field $\eta^{A}$ transform like
\beq
\delta \eta^{A}=\alpha_{A}m_{A}+\text{linear\,terms} \,.
\eeq
Finally, the effective operators in \eq{eq:Ltoyeff} 
written in terms of the $Z_{\mu}^{A}$ and $\eta^{A}$ fields are 
\beq
\sum_{A,B,C}\frac{\tilde{g}_{A}\tilde{g}_{B}\tilde{g}_{C}}{48\pi^{2}}C_{Z}^{ABC}\epsilon^{\alpha\mu\nu\beta}Z_{\alpha}^{A}Z_{\mu}^{B}\partial_{\beta}Z_{\nu}^{C}-\sum_{A,B,C}\frac{\tilde{g}_{A}\tilde{g}_{B}\tilde{g}_{C}}{48\pi^{2}}D_{\eta}^{ABC}\epsilon^{\mu\nu\alpha\beta}\frac{\eta^{A}}{m_{A}}\partial_{\alpha}Z_{\mu}^{B}\partial_{\beta}Z_{\nu}^{C}+\dots,
\eeq
where the dots contain the interaction terms with the Higgs physical modes.
The rotated EFT coefficients are equal to
\beq
\label{Cz}
\begin{split}
C_{Z}^{ABC}=\int_{0}^{+\infty}\!\!\!\text{d}s\!\int_{0}^{1}\!\text{d}x\!&\int_{0}^{1}\!\text{d}y\!\int_{0}^{1}\!\text{d}z\,2\,\delta(1\!-\!x\!-\!y\!-\!z)\times\\
\times\,\text{Re}\Biggl\{ 3y\,&\text{Tr}\!\left[e^{-sy\mathcal{M}^{\dag}\mathcal{M}}T_{R}^{A}\mathcal{M}^{\dag}e^{-sz\mathcal{M}\mathcal{M}^{\dag}}T_{L}^{B}e^{-sx\mathcal{M}\mathcal{M}^{\dag}}T_{L}^{C}\mathcal{M}\right]\\
-3y\,&\text{Tr}\!\left[e^{-sy\mathcal{M}^{\dag}\mathcal{M}}T_{R}^{B}\mathcal{M}^{\dag}e^{-sz\mathcal{M}\mathcal{M}^{\dag}}T_{L}^{A}e^{-sx\mathcal{M}\mathcal{M}^{\dag}}T_{L}^{C}\mathcal{M}\right]\\
+3y\,&\text{Tr}\!\left[e^{-sy\mathcal{M}\mathcal{M}^{\dag}}T_{L}^{B}\mathcal{M}e^{-sz\mathcal{M}^{\dag}\mathcal{M}}T_{R}^{A}e^{-sx\mathcal{M}^{\dag}\mathcal{M}}T_{R}^{C}\mathcal{M}^{\dag}\right]\\
-3y\,&\text{Tr}\!\left[e^{-sy\mathcal{M}\mathcal{M}^{\dag}}T_{L}^{A}\mathcal{M}e^{-sz\mathcal{M}^{\dag}\mathcal{M}}T_{R}^{B}e^{-sx\mathcal{M}^{\dag}\mathcal{M}}T_{R}^{C}\mathcal{M}^{\dag}\right]\\
+y\,&\text{Tr}\!\left[e^{-sy\mathcal{M}^{\dag}\mathcal{M}}\mathcal{M}^{\dag}\mathcal{M}T_{R}^{A}e^{-sz\mathcal{M}^{\dag}\mathcal{M}}T_{R}^{B}e^{-sx\mathcal{M}^{\dag}\mathcal{M}}T_{R}^{C}\right]\\
-x\,&\text{Tr}\!\left[e^{-sy\mathcal{M}^{\dag}\mathcal{M}}T_{R}^{A}e^{-sz\mathcal{M}^{\dag}\mathcal{M}}T_{R}^{B}e^{-sx\mathcal{M}^{\dag}\mathcal{M}}\mathcal{M}^{\dag}\mathcal{M}T_{R}^{C}\right]\\
+x\,&\text{Tr}\!\left[e^{-sy\mathcal{M}\mathcal{M}^{\dag}}T_{L}^{A}e^{-sz\mathcal{M}\mathcal{M}^{\dag}}T_{L}^{B}e^{-sx\mathcal{M}\mathcal{M}^{\dag}}\mathcal{M}\mathcal{M}^{\dag}T_{L}^{C}\right]\\
-y\,&\text{Tr}\!\left[e^{-sy\mathcal{M}\mathcal{M}^{\dag}}\mathcal{M}\mathcal{M}^{\dag}T_{L}^{A}e^{-sz\mathcal{M}\mathcal{M}^{\dag}}T_{L}^{B}e^{-sx\mathcal{M}\mathcal{M}^{\dag}}T_{L}^{C}\right]\Biggr\} \, ,
\end{split}
\eeq
and
\beq
\label{Deta}
\begin{split}
D_{\eta}^{ABC}=\int_{0}^{+\infty}\!\!\!\!\!\!\text{d}s\!\int_{0}^{1}\!\text{d}x\!\int_{0}^{1}\!&\text{d}y\!\int_{0}^{1}\!\text{d}z\,6\,\delta(1\!-\!x\!-\!y\!-\!z)\times\\
\times\,\text{Re}\Biggl\{x\,&\text{Tr}\!\left[e^{-sz\mathcal{M}^{\dag}\mathcal{M}}T_{R}^{B}\mathcal{M}^{\dag}e^{-sx\mathcal{M}\mathcal{M}^{\dag}}T_{L}^{C}e^{-sy\mathcal{M}\mathcal{M}^{\dag}}(T_{L}^{A}\mathcal{M}-\mathcal{M}T_{R}^{A})\right]\\
+x\,&\text{Tr}\!\left[e^{-sz\mathcal{M}^{\dag}\mathcal{M}}T_{R}^{C}\mathcal{M}^{\dag}e^{-sx\mathcal{M}\mathcal{M}^{\dag}}T_{L}^{B}e^{-sy\mathcal{M}\mathcal{M}^{\dag}}(T_{L}^{A}\mathcal{M}-\mathcal{M}T_{R}^{A})\right]\\
+y\,&\text{Tr}\!\left[e^{-sy\mathcal{M}^{\dag}\mathcal{M}}\mathcal{M}^{\dag}T_{L}^{C}e^{-sx\mathcal{M}\mathcal{M}^{\dag}}T_{L}^{B}e^{-sz\mathcal{M}\mathcal{M}^{\dag}}(T_{L}^{A}\mathcal{M}-\mathcal{M}T_{R}^{A})\right]\\
+y\,&\text{Tr}\!\left[e^{-sy\mathcal{M}^{\dag}\mathcal{M}}\mathcal{M}^{\dag}T_{L}^{B}e^{-sx\mathcal{M}\mathcal{M}^{\dag}}T_{L}^{C}e^{-sz\mathcal{M}\mathcal{M}^{\dag}}(T_{L}^{A}\mathcal{M}-\mathcal{M}T_{R}^{A})\right]\\
+y\,&\text{Tr}\!\left[e^{-sz\mathcal{M}^{\dag}\mathcal{M}}T_{R}^{B}e^{-sx\mathcal{M}^{\dag}\mathcal{M}}T_{R}^{C}\mathcal{M}^{\dag}e^{-sy\mathcal{M}\mathcal{M}^{\dag}}(T_{L}^{A}\mathcal{M}-\mathcal{M}T_{R}^{A})\right]\\
+y\,&\text{Tr}\!\left[e^{-sz\mathcal{M}^{\dag}\mathcal{M}}T_{R}^{C}e^{-sx\mathcal{M}^{\dag}\mathcal{M}}T_{R}^{B}\mathcal{M}^{\dag}e^{-sy\mathcal{M}\mathcal{M}^{\dag}}(T_{L}^{A}\mathcal{M}-\mathcal{M}T_{R}^{A})\right]\Biggr\}\,.
\end{split}
\eeq

\subsection{Properties of the WZ coefficients}

In general, the expressions \eqref{Cz} and \eqref{Deta} involve non-trivial integrations which are difficult to compute. Special simplifications occur if the fermion mass term $\bar\psi_{Li}\mathcal{M}_{ij}\psi_{Rj}$ is invariant under any of the symmetry generators $T^{A}$. If so, the mass matrix $\mathcal{M}$ satisfies
\beq
\label{inv1}
\sum_{k}\mathcal{M}_{ik}(T_{R}^{A})_{kj}-\sum_{k}(T_{L}^{A})_{ik}\mathcal{M}_{kj}=0 \,.
\eeq
Alternatively, the invariance of the mass term reads
\beq
\sum_{a,b}\mathcal{Y}_{ij}^{a}(T_{H}^{A})_{ab}v_{b}=0 \,,
\eeq
which could occur if $T^{A}$ belongs to $\mathcal{G}_{0}$ or some Yukawa coupling vanishes. Then, one finds
\beq
\label{simplified}
C_{Z}^{ABC}=\begin{cases}
\text{Tr}\,T_{R}^{A}\{T_{R}^{B},T_{R}^{C}\}-\text{Tr}\,T_{L}^{A}\{T_{L}^{B},T_{L}^{C}\} & \text{if \quad $\sum_{a,b}\mathcal{Y}_{ij}^{a}(T_{H}^{B,C})_{ab}v_{b}=0$\, ,}\\
\text{Tr}\,T_{L}^{A}\{T_{L}^{B},T_{L}^{C}\}-\text{Tr}\,T_{R}^{A}\{T_{R}^{B},T_{R}^{C}\} & \text{if \quad $\sum_{a,b}\mathcal{Y}_{ij}^{a}(T_{H}^{A,C})_{ab}v_{b}=0$\, ,}\\
0 & \text{if \quad $\sum_{a,b}\mathcal{Y}_{ij}^{a}(T_{H}^{A,B})_{ab}v_{b}=0$\, .}
\end{cases}
\eeq
and
\beq
\label{simplified2}
D_{\eta}^{ABC}=3\left[ \text{Tr}\,T_{L}^{A}\{T_{L}^{B},T_{L}^{C}\}-\text{Tr}\,T_{R}^{A}\{T_{R}^{B},T_{R}^{C}\} \right]\,\,\,\,\,\,\,\,\,\,\,\text{if \quad $\sum_{a,b}\mathcal{Y}_{ij}^{a}(T_{H}^{B,C})_{ab}v_{b}=0$}\,.
\eeq
Note that $D_{\eta}^{ABC}$ vanishes if
\beq
\sum_{a,b}\mathcal{Y}_{ij}^{a}(T_{H}^{A})_{ab}v_{b}=0\,,
\eeq
i.e.~if the fermion mass term is invariant under symmetry generator $T^{A}$.\\
\\
Consider now an unbroken generator $Q\in\mathcal{G}_{0}$, hence satisfying
\beq
\label{inv2}
\sum_{k}\mathcal{M}_{ik}(Q_{R})_{kj}-\sum_{k}(Q_{L})_{ik}\mathcal{M}_{kj}=0\,,
\eeq
with the commutation rules
\beq
[Q,T^{A}]=\sum_{B}q_{AB}T^{B}\,.
\eeq
Thanks to \eqref{inv2} and the cyclic property of the trace, the expression
\beq
C_{Z}^{ABC}|_{T^{A} \rightarrow [Q,T^{A}]}\,+\,C_{Z}^{ABC}|_{T^{B} \rightarrow [Q,T^{B}]}\,+\,C_{Z}^{ABC}|_{T^{C} \rightarrow [Q,T^{C}]}\,=0
\eeq
is identically zero. Then, we find that the EFT coefficients satisfy
\beq
\sum_{D}(q_{AD}C_{Z}^{DBC}+q_{BD}C_{Z}^{ADC}+q_{CD}C_{Z}^{ABD})=0\,.
\eeq
The same argument yields
\beq
\sum_{D}(q_{AD}D_{\eta}^{DBC}+q_{BD}D_{\eta}^{ADC}+q_{CD}D_{\eta}^{ABD})=0 \,.
\eeq

\begin{small}

\bibliographystyle{utphys}
\bibliography{bibliography}

\end{small}

\end{document}